\documentclass[prx,twocolumn,superscriptaddress]{revtex4-2}
\usepackage[utf8]{inputenc}
\usepackage{xcolor}
\usepackage{graphicx, graphics,epsfig}
\usepackage{amssymb,amsfonts,amsmath}
\usepackage{ifthen}	

\usepackage{graphics,graphicx,epsfig}
\usepackage{epsf,epstopdf,wrapfig}
\usepackage{amssymb,amsfonts,amsmath}
\usepackage[export]{adjustbox}
\usepackage[english]{babel}
\include{epsf}
\usepackage{ifthen}

\usepackage{graphics,graphicx,epsfig}
\usepackage{amssymb,amsfonts,amsmath}
\usepackage{bbm}
\usepackage{ifthen}
\usepackage{color}
\usepackage[colorlinks = true,
	     linkcolor = cyan,
            urlcolor  = cyan,
            citecolor = cyan,
            anchorcolor = black]{hyperref}
\graphicspath{{Figures/}}	
\usepackage{rotating}
\usepackage{csvsimple}
\usepackage{nicematrix}
\usepackage{cancel}
\usepackage{comment}

\usepackage{color}   
\usepackage[normalem]{ulem}

\newcommand{\EQ}{\begin{equation}}
\newcommand{\EE}{\end{equation}}
\newcommand{\EQA}{\begin{eqnarray}}
\newcommand{\EEA}{\end{eqnarray}}

\newcommand{\st}{{\text{st}}}

\newcommand{\U}{\mathbf{U}} 
\newcommand{\BB}{\mathbf{B}}

\renewcommand{\d}{{\text{d}}}

\usepackage[normalem]{ulem}

\newcommand{\UI}{\cancel{\mathrm{\bf U}}}

\newcommand{\corres}{Correspondence should be addressed to:
Armita Nourmohammad: {armita@uw.com}.}
\newcommand{\cofirst}{These authors contributed equally.}

\begin{document}

\title{Adaptive efficiency of information processing in immune-pathogen co-evolution}
\author{Quinn N Bellamy}
\thanks{\cofirst}
\affiliation{Department of Physics, University of Washington, 3910 15th Avenue Northeast, Seattle, WA 98195, USA}
\author{Zachary Montague}
\thanks{\cofirst}
\affiliation{Department of Physics, University of Washington, 3910 15th Avenue Northeast, Seattle, WA 98195, USA}
\author{Luca Peliti}
\affiliation{Santa Marinella Research Institute,  00058 Santa Marinella, Italy}
\author{Armita Nourmohammad}
\thanks{\corres}
 \affiliation{Department of Physics, University of Washington, 3910 15th Avenue Northeast, Seattle, WA 98195, USA}
\affiliation{Paul G. Allen School of Computer Science and Engineering, University of Washington, 85 E Stevens Way NE, Seattle, WA 98195, USA}
\affiliation{Department of Applied Mathematics, University of Washington, 4182 W Stevens Way NE, Seattle, WA 98105, USA}
 \affiliation{Fred Hutchinson Cancer Center, 1241 Eastlake Ave E, Seattle, WA 98102, USA}

\begin{abstract}
Organisms have evolved immune systems that can counter pathogenic threats. The adaptive immune system in vertebrates consists of  a diverse repertoire of immune receptors that can dynamically reorganize to specifically target the ever-changing pathogenic landscape. Pathogens in return  evolve to escape the immune challenge, forming an  co-evolutionary arms race. We introduce a formalism to characterize out-of-equilibrium interactions in co-evolutionary processes. We show that the rates of information exchange and entropy production can distinguish the leader from the follower in an evolutionary arms races.  Lastly, we introduce co-evolutionary  efficiency as a metric to quantify each population's ability to exploit information in response to the other. Our formalism provides insights into the conditions necessary for stable co-evolution and establishes bounds on the limits of information exchange and adaptation in co-evolving systems.
\end{abstract}
\maketitle

\section{Introduction} 
Organisms, ranging from bacteria to humans, possess immune systems that have evolved to counteract pathogenic threats in their environments. In vertebrates, including humans, the adaptive immune system mounts a flexible and diverse response to neutralize pathogens.  This system is shaped by eco-evolutionary dynamics of immune cells during the host's lifetime, as it detects infections and generates and reorganizes its diverse repertoire of immune cells to specifically target  and clear pathogens~\cite{Janeway2001-ws}. The CRISPR-Cas immunity in bacteria operates on the same principle to combat infections by phages, which are viruses that attack bacteria~\cite{Hampton2020-ds,Bradde2020-iy,Watson2021-ul}. On the other hand, pathogens, especially viruses, thrive by transmission through multiple hosts and evolve to escape the hosts’ immune defenses over time. In response, the host updates its immune repertoire to battle the escaping pathogens, resulting in an out-of-equilibrium co-evolutionary arms race~\cite{Nourmohammad2016-zg}.

Focusing on vertebrates, immune-pathogen co-evolution occurs on multiple scales. Intra-host co-evolution, for example, can shape infections with chronic viruses like HIV~\cite{Liao2013-ga,Luo2015-mj,Wang2015-rf,Nourmohammad2016-zg,Nourmohammad2019-lk, Strauli2019-xx,Mazzolini2023-sd}. In this case, the immune B-cell receptors (BCRs) undergo rapid Darwinian somatic evolution through affinity maturation to  accumulate beneficial mutations that improve their affinities for binding to an infecting viral variant~\cite{Victora2022-an}. Viruses then evolve within each host to escape this immunity, prompting further affinity maturation of B-cell receptors. At the population level, viruses transmit from one host to another eliciting independent immune responses across individuals. Viruses evolve to escape the resulting population-level immunity, again driving a co-evolutionary arms race, apparent in viral escape during the COVID pandemic or in the evolution of the seasonal influenza virus~\cite{Barrat-Charlaix2024-gx,Meijers2023-sr}.
 
In both the intra-host and population-level co-evolutionary processes, the affinity of interactions between immune receptors and viral epitopes is a key molecular phenotype that determines the (evolutionary) fitness of either side. Pathogens escape through epitope mutations that reduce their affinity while immune receptors are selected for their heightened affinities. The map from genotypes to a given phenotype describing molecular function (e.g., protein binding affinity) is highly degenerate, and so is the map from combinations of phenotypes to fitness, as a measure of the reproductive success of an individual. Population genetics models have been developed to study the effects of mutations, selection, and stochasticity on such co-evolutionary processes in both the genotype and the phenotype spaces~\cite{Nourmohammad2016-zg,Marchi2021-xd,  Schnaack2021-fo,Schnaack2022-if, Chardes2023-if, Meijers2023-sr, Barrat-Charlaix2024-gx}. The picture that emerges from all these descriptions has a lot in common with how a statistical description of  physical systems with large number of degrees of freedom, e.g. gas molecules,  can be mapped into a thermodynamic model for macroscopic quantities, such as pressure and energy~\cite{Barton2009-ak,Nourmohammad2013-hs,Nourmohammad2013-tm}.  This connection has been explored extensively for populations in equilibrium~\cite{Sella2005-br,Mustonen2005-yb}, and more formal analogies have been drawn between stochastic thermodynamics and driven evolutionary processes in time-dependent environments~\cite{Mustonen2009-uh, Mustonen2010-fj, Held2014-tn,PhysRevLett.115.238102,PhysRevE.95.012131, Nourmohammad2017-vr}.  Although compelling, these analogies are not exact. Notably, evolutionary theory is  a stochastic process and does not have a conserved quantity resembling the energy in thermodynamics (i.e., it does not satisfy a conservation law similar to the first law of thermodynamics). However, there is a strong correspondence between the second law of thermodynamics concerning entropy production in physical systems and the rules governing the production of information theoretical entropy for evolving populations. Notably, the fitness flux theorem was developed~\cite{Mustonen2010-fj} in close analogy with the concept of entropy production in stochastic thermodynamics~\cite{Jarzynski1997-ia, Crooks1999-he,Seifert2012-dg}; fitness flux quantifies the extent of out-of-equilibrium adaptation a population undergoes in response to changing environments.

In this work, we focus on co-evolutionary processes, in which the feedback between the two populations---and not an external time-varying environment---is the source of the evolutionary drive. We propose a  bipartite discrete-time Markov process for interacting subsystems, as a proxy to model immune-pathogen interactions. While this approach comes at the cost of biological over-simplification, it ensures the tractability and interpretability of our models, offering clear insights into the newly introduced thermodynamics concepts in biological co-evolution. Specifically, we introduce the rates of information exchange and entropy production as a way to identify the leader and the follower in a co-evolutionary system. We introduce the concept of co-evolutionary efficiency, which generically applies to stochastic processes that are not constrained by the first law of thermodynamics (i.e., in which there is no conserved quantity like energy). Co-evolutionary  efficiency quantifies the efficacy of each population in exploiting information in response to the other population. It defines conditions for stable co-evolution, and places bounds on the limits of information exchange and adaptation in co-evolving systems.

\section{Model}
\begin{figure*}[t!]
\includegraphics[width=0.7\textwidth]{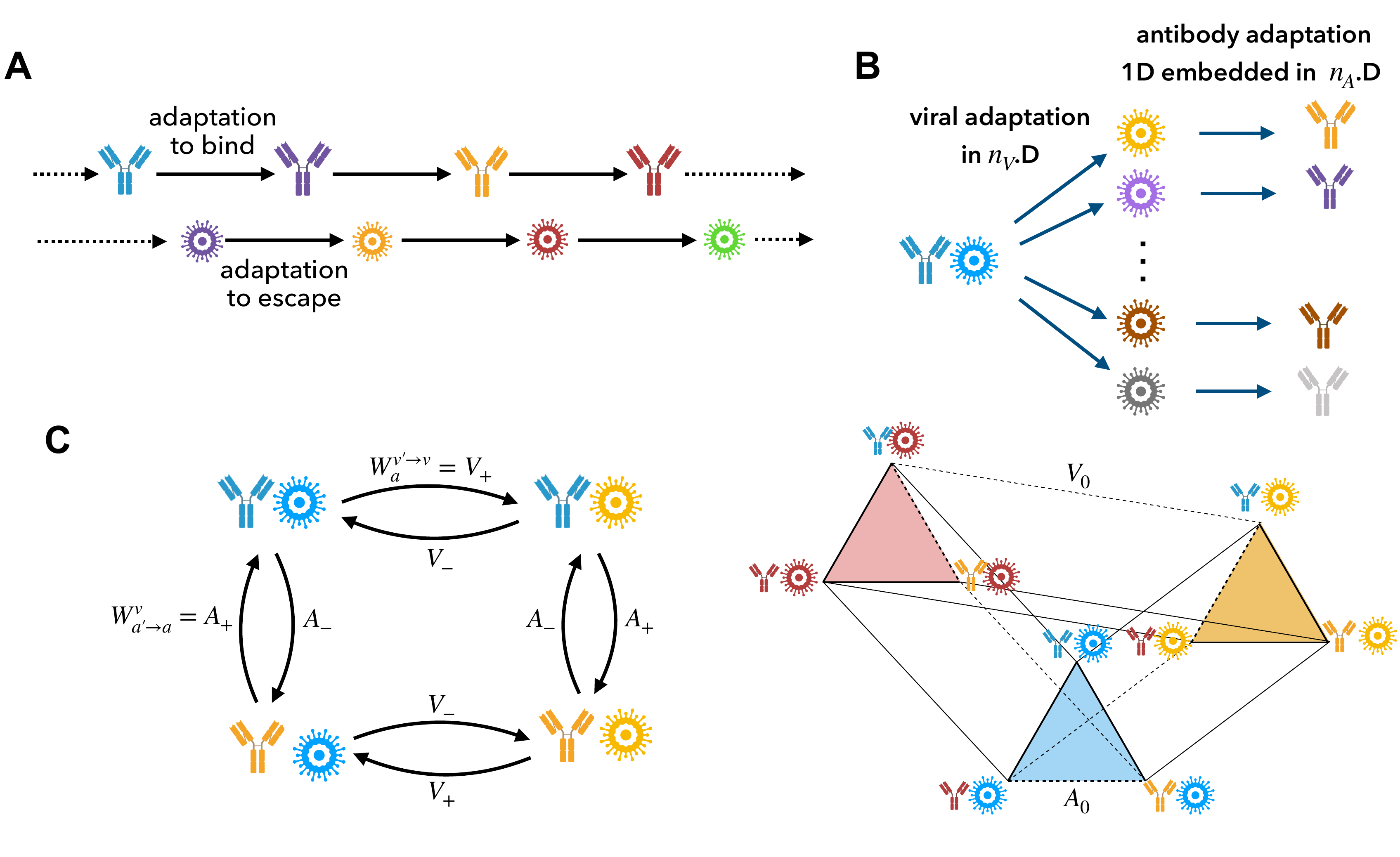}
\caption{{\bf Bipartite stochastic model of immune-pathogen co-evolution.} {\bf (A)} Antibodies adapt to bind to the circulating viruses (match the color), while viruses adapt to escape the antibodies (mismatch the color). {\bf (B)} We assume that viruses can escape to  $n_V-1$ states (out of $n_V$) that do not match the color of the circulating antibody. To  bind to and neutralize a given virus, antibodies have only a single choice to evolve to match the virus (i.e., the state with the same color). As a result, the beneficial substitution of antibodies to bind to a given virus happens along 1 dimension, embedded in an $n_A$ dimensional space of possible antibodies.  For most of our analyses, we assume a balanced evolutionary space where $n_A= n_V =n$. {\bf (C)} We assume that  the system is bipartite, whereby transitions happen exclusively in one or the other population, but never simultaneously (eq.~\eqref{eq.W_bipartite}). The bipartite graph for $n=2$ (left) and $n=3$ (right) are shown. We assume that the transition probabilities between different states depend on whether the new state is beneficial or deleterious for the transitioning subpopulation. For antibodies, beneficial transitions with probability $W_{a'\to a}^v = A_+$  result in a matching color with the circulating virus ($a=v$), and deleterious transitions  with probability $W_{a'\to a}^v = A_-$  result in a mismatched color ($a\neq v$). For $n\geq 3$, neutral transitions can occur, whereby the system remains unbound after the transition. Neutral transitions occur with probability $A_0$ for antibodies, and with probability $V_0$ for viruses, and are indicated by dashed lines for the case with $n=3$ (right). Beneficial and deleterious transitions are indicated by full lines.
 }
\label{fig:bipartite}
\end{figure*}

We assume  interacting adversarial populations of  immune agents and pathogens, which without loss of generality, we refer to  as antibodies and viruses, respectively. We assume that the dynamics of the joint system follows a  Markov process, whereby the state of the joint system at time $t$ only depends on its state at time $t-1$ and not on its prior history (Fig.~\ref{fig:bipartite}A).

To describe this process more concretely, let $A$ represent the antibody ensemble and $V$  the viral ensemble. We represent  the space of realizable states of $A$ by $\Omega(A) = \{1,\dots, n_A\}$ and the space of realizable states of $V$ by $\Omega(V) = \{1,\dots,  n_V \}$, where $n_A$ and $n_V$ are the dimensionality of the antibody and virus ensemble respectively. We represent a joint state of the antibody-virus system as $(a, v)$, where $a \in \Omega(A)$ and $v \in \Omega(V)$. For instance, for the case of $n_A = n_V = 2$, all the possible states $(a,v)$ that can be realized are $\{(1, 1), \,(1, 2), \,(2, 1), \,(2, 2)\}$.

We describe the stochastic transitions between two arbitrary (joint) states $(a',v') \to (a,v)$ by an $N \times N$  discrete-time Markov transition matrix $W$, where $N = n_A n_V$ is the number of possible joint states. Each entry of $W$ represents the probability of transitioning from one state to another. Specifically, we use the convention in which the matrix element $W_{ji} = W_{a'\to a}^{v'\to v}$ denotes the transition probability from the state $i \equiv (a',v')$ (column)  to the state $j\equiv (a,v)$ (row). The diagonal elements of the matrix are set by normalization: $W_{a\to a}^{v\to v} =1- \sum_{(a',v')\neq (a,v)}W_{a\to a'}^{v\to v'}$.  Moreover, to make the problem analytically tractable, we  assume that the system is bipartite, whereby transitions happen exclusively in one or the other population, but never simultaneously. Analogous bipartite treatments have been done for the analysis of information exchange between physical systems within the framework of stochastic thermodynamics~\cite{Sagawa2012-ti, Sagawa2013-in, Barato2014-al,Horowitz2014-od}. The resulting bipartite transition matrix is given by
\EQ
W_{a'\to a}^{v'\to v} =
\begin{cases}
    W_{a'\to a}^{v} & \text{if}\,  v=v', \\
    W_{a}^{v'\to v} & \text{if} \, a=a', \\
    0 & \text{otherwise.}
\end{cases}
\label{eq.W_bipartite}
\EE
The fitness of antibodies or viruses depend on their binding abilities: a bound antibody-virus complex can lead to the neutralization of the virus, while an unbound virus is able to evade immunity, causing more infections. In our model, we assume that a state $(i,j)$ is bound if $i=j$, and unbound otherwise (Fig.~\ref{fig:bipartite}B). With this definition, we implicitly disregard cross-reactivity in interactions, which refers to the ability of a given antibody (or virus) to react with targets other than its cognate one, often with slightly reduced affinity. 

Our model can allow for an imbalanced number of viral and antibody states, i.e.,  $n_V \neq n_A$. In the case that $n_V>n_A$ there exists states that are inaccessible to the antibodies (i.e., $v=n_A+1,\dots,n_V$) and if the virus transitions to any of them, no change in the antibody state can result in binding. As a result, the virus can simply hide in the inaccessible states, leading to the extinction of the host---and the breakdown of the co-evolutionary dynamics---as no immune response can counter infections. On the other hand, if  $n_A>n_V$, for each viral state there exists more antibody states that do not bind to the virus, compared to the balanced case of $n_A=n_V$. This  results in an evolutionary distraction for antibodies, which reduces the efficiency of the antibody response to viruses but does not qualitatively impact the co-evolutionary process.  Although mathematically interesting, the imbalanced case $n_A\neq n_V$ does not offer additional biological insight into the co-evolutionary dynamics. Therefore, for the clarity of our derivations, we will  consider the balanced case of $n_A=n_V=n$ in the main text and only offer discussions about the imbalanced case in Appendix~\ref{Appendix-imbalance}. 

In our bipartite model, we assume that the transition probabilities between different states in each subpopulation only depend on whether the ensuing phenotypic change is beneficial or deleterious for the evolving (transitioning) subpopulation. 
As such, we set that for the antibodies,  beneficial changes (transitions from unbound to bound) occur with a probability $A_+$, deleterious changes (transitions from bound to unbound) occur with a probability $A_-$, and neutral changes that do not modify the binding phenotype of the antibody occur with a probability $A_0$ (Fig.~\ref{fig:bipartite}C). Similarly, for  viruses, we assume that the beneficial changes (transitions from bound to unbound) occur with a probability $V_+$, deleterious changes (transitions from unbound to bound) occur with a probability $V_-$, and neutral changes occur with a probability $V_0$ (Fig.~\ref{fig:bipartite}C). The resulting non-zero off-diagonal elements of the transition matrix $W$ can be expressed as
\EQA
\nonumber W_{a'\to a}^{v} &=& \underbrace{(1- \delta_{a',v})\delta_{a,v}}_{\text{unbound $\to$ bound}} \,A_+ + \underbrace{ \delta_{a',v}(1-\delta_{a,v})}_\text{bound $\to$ unbound} \, A_- \\
  &&+ \underbrace{\left[\delta_{a',v}\delta_{a,v}+ (1-\delta_{a',v})(1-\delta_{a,v}) \right]}_\text{no phenotypic change}\, A_0,\\
\nonumber W_{ a}^{v'\to v} &=& \underbrace{\delta_{a,v'}(1-\delta_{a,v})}_{\text{bound $\to$ unbound}} \,V_+ + \underbrace{ (1-\delta_{a,v'})\delta_{a,v}}_\text{unbound $\to$ bound} \, V_- \\
  &&+ \underbrace{\left[\delta_{a,v'}\delta_{a,v}+ (1-\delta_{a,v'})(1-\delta_{a,v}) \right]}_\text{no phenotypic change}\, V_0.
\label{eq.W}
\EEA
where $\delta_{x,y}$ is the Kronecker delta function, taking the value $1$ when $x=y$, and 0, otherwise. These transition probabilities should be interpreted as substitutions in a population, which is the probability for a variant carrying a specific mutation to grow and take over the population (i.e., fix) by natural selection. Even though the baseline mutation rates should not depend on the phenotypic effect of a mutation, the substitution rates do. Beneficial mutations are exponentially more likely to fix in a population than their deleterious counterparts~\cite{Kimura1968-nb}. As such, we assume that $A_+>A_0>A_-$, and $V_+>V_0 >V_-$.  Lastly, our model implies that both  the antibody and viral  populations are monomorphic, meaning that they are comprised of a single type at a given time, which can be substituted by another type by the transition probability $W$. A more realistic evolutionary model should include the effect of diversity and  polymorphism in each population.

The bound and unbound states form the two relevant phenotypes for evolutionary transitions. For the case with a balanced number of antibody and viral states ($n_A=n_V= n$), we can argue by symmetry that the diagonal  terms of the transition matrix $W_{a\to a}^{v\to v}$ fall  into two classes: the ones associated with  bound states  ($a=v$), which we denote by $W_{\bf B}$, and the ones associated with unbound states $a\neq v$, which we denote by $W_{\bf U}$.  
These diagonal elements are the probabilities that the system remains in the state that it starts at. These two probabilities are given by
\EQA
 W_{\bf B} &=& 1-  (n-1)\left(V_++A_-\right),\label{eq:WB_diagonal}\\
  W_{\bf U} &=&1- \left[ V_-+A_+ + (n-2)(A_0+V_0)\right].\label{eq:WU_diagonal}
\EEA
The diagonal element in each case is the complement of the probability to leave the specified state.  For example, the probability of remaining in a bound state $W_\BB$ is the complement of the sum of  probabilities for all the transitions out of a bound state. Since our model allows for only one bound state for a given antibody and virus, i.e., the state $a=v$, all  transitions from a bound state would end in an unbound configuration. There are $n-1$ such transitions that can be reached either by beneficial changes in viruses or deleterious changes in antibodies, with substitution probabilities $V_+$ and $A_-$, respectively. Therefore, $W_\BB$ can be expressed by eq.~\eqref{eq:WB_diagonal}. Similarly,   $W_\U$ in eq.~\eqref{eq:WU_diagonal} is evaluated by accounting for all the possible transitions out of an unbound state, i.e., the transition to a single bound state by either antibodies or viruses, and the  neutral transitions to any of the other $n-2$ unbound states (with rates  $A_0$ and $V_0$ for antibodies and viruses);  see Appendix~\ref{Appendix-imbalance} for a discussion on the imbalanced case. 

Interestingly, the fact that the probabilities of remaining in a given state (diagonal elements) must be between zero and one sets an upper bound for the beneficial substitution probability of  viruses that strongly depend on the dimensionality of the system (from eq.~\ref{eq:WB_diagonal}),
\EQ V_+ \leq \frac{1- (n-1)A_-}{n-1}
\label{eq.boundVplus}
\EE
 Similarly, eq.~\ref{eq:WU_diagonal} sets the upper bound for the   beneficial substitution probability of  antibodies  as $A_+ \leq 1- (V_- + (n-2) (A_0+V_0))$. The neutral substitution probabilities, $A_0$ and $V_0$, scale as $1/N_\text{eff}$  with the effective population size of each system $N_\text{eff}$~\cite{Kimura1968-nb}, which is likely to be much larger than the dimensionality of the system $n$, which follows $ (n-2) (A_0+V_0)\ll 1$. Therefore, given that the deleterious substitution probabilities are exponentially small $V_-\ll 1$, the upper bound on the beneficial substitution probability for antibodies can be approximated as $A_+\lesssim 1$, with no strong dependence on the dimensionality of the system.

\section{Results}
\subsection{Stationary state statistics}
\noindent {\bf \small Stationary state probabilities.} The stationary state probability of all the possible $n^2$ states, which we denote by the vector $P_\st$, should follow from $P_\st= W P_\st $, i.e., $P_\st$ is the eigenvector of the transition matrix associated with eigenvalue $1$. To solve this eigenvector problem, let us denote the stationary state probability associated with the bound state $(a,v)$ by $P_\st(a,v;\BB)$, which can be expressed as
\EQA
\nonumber  P_\st(a,v;\BB)&=& W_{\bf B} P_\st(a,v; \BB) +\sum_{a'\neq a,v} W_{a'\to a}^v P_\st(a',v; \U) \\
 &&+  \sum_{v'\neq v,a} W_a^{v'\to v} \, P_\st(a,v'; \U) .\label{eq.Pstat0}
\EEA
The first term accounts for remaining in the bound $(a,v)$ state with probability $W_\BB$ (diagonal term from eq.~\eqref{eq:WB_diagonal}). The other term accounts for the transitions from all the accessible $n-1$ unbound states to the bound state $(a,v)$ through either beneficial antibody substitutions with probabilities  $W_{a'\to a}^v =A_+$ (second term), or deleterious viral substitutions with probabilities $W_a^{v'\to v} = V_-$ (third term). By symmetry, we can assume that all the $n$ bound states ($\forall a = v$) have the same probability $P_\st(a,v; \BB)\equiv\pi_\BB$, and all the $n(n-1)$ unbound states ($\forall a'\neq v$) have the same probability $ P_\st(a',v; \U)\equiv\pi_\U$. We can thus simplify eq.~\eqref{eq.Pstat0} and, by requiring the probabilities to be normalized, arrive at the expressions for the stationary state probabilities associated with being in a given bound or an unbound state,
\EQA
 \pi_\BB = \frac{A_+ + V_-}{Z}, \qquad\qquad \pi_\U=\frac{A_- + V_+}{Z},
\label{eq.Pstat1}
\EEA 
where $Z= n (A_+ + V_-) + n(n-1) (A_- + V_+)$ is the normalization constant.\\
\\
\noindent{\bf \small Probability current}. The probability current $J_{a'\to a}^{v'\to v}$ between states $(a',v') \to (a,v)$ quantifies the change in the probability density:
 \EQA 
  \frac{\d}{\d t}P (a,v)  &\equiv & \sum_{a',v'} J_{a'\to a}^{v'\to v}\label{eq:prob_current} \\
\nonumber &=& \sum_{a',v'} P(a',v') W_{a'\to a}^{v'\to v} - P({a,v}) W_{a\to a'}^{v\to v' } .
\EEA
Although in the stationary state (i.e., when ${\d_tP (a,v)=0}$), the total probability current to any state is zero, the individual elements of the probability current  matrix $J_{a'\to a}^{v'\to v}$ can still be non-zero, consistent with the fact that the system is out of equilibrium, i.e., it does not satisfy detailed balance. In our bipartite system, the probability current splits into partial components  associated with changes in the antibodies $J_{a'\to a}^{v} = P({a',v}) W_{a'\to a}^{v } -P({a,v}) W_{a\to a'}^{v } $  and in the viral subsystem $J_{a}^{v'\to v} = P({a,v'}) W_{a}^{v'\to v} - P({a,v})
 W_{a}^{v\to v'}$, conditioned on the state of the other subsystem. 
 
 By inspection, one can see that the elements of the  probability current  (and the conditional components) are asymmetric, i.e.,   $J_{a'\to a}^{v'\to v} = - J_{a\to a'}^{v\to v'}$, and  their magnitudes depend on the phenotypic effect of the transition. Therefore, given the bipartite nature of the system, the problem of computing the probability current elements reduces to computing the favorable and the neutral conditional currents;  the currents associated with the unfavorable transitions are the negative of the inverted favorable transitions. Using the definition of probability current in eq.~\eqref{eq:prob_current}, we can evaluate the  conditional currents associated with (i) favorable changes in antibodies from the unbound state $(a',v)$ to the bound state $(a,v)$, i.e.,  $J_{a':\U\to a:\BB}^v$,  (ii) favorable  changes in viruses from a bound to an unbound state, i.e.,  $J_{ a}^{v':\BB\to v:\U}$, and (iii) neutral changes in antibodies and viruses, denoted by  $J_{a':\U\to a:\U}^v$ and  $J_{ a}^{v':\U\to v:\U}$, respectively.  The   elements of the  probability current are thus given by   
\EQA
\nonumber J_{a':\U\to a:\BB}^v  = - J_{a:\BB\to a':\U}^v  &=& \pi_\U A_+  - \pi_\BB A_- =  \frac{\sigma}{Z}, \\
\nonumber J_{ a}^{v':\BB\to v:\U} = - J_{a}^{v:\U\to v':\BB} &=& \pi_\BB V_+  - \pi_\U V_- =  \frac{\sigma}{Z}, \\
 J_{a':\U\to a:\U}^v = J_{ a}^{v':\U\to v:\U}  &=&0,
\label{eq:J-Phenotype}
\EEA
where $\sigma = A_+V_+-A_-V_-$, and $Z$ is the normalization constant from eq.~\eqref{eq.Pstat1}. The probability current matrix takes a simple form in which the elements associated with favorable, unfavorable, and neutral transitions are $\sigma/Z$, $-\sigma/Z$, and $0$, respectively.   

\subsection{Fitness and transfer flux as rates of entropy production and information exchange in  co-evolving systems}
The two quantities of fitness and transfer flux have been previously introduced to measure the extent of non-equilibrium drive in evolving  populations~\cite{Mustonen2009-uh, Mustonen2010-fj,Nourmohammad2016-zg}.  To characterize the out-of-equilibrium drive in our bipartite co-evolutionary system, we start with fitness flux, which measures the rate of adaptation in a population. Fitness flux was introduced in ref.~\cite{Mustonen2010-fj} in analogy to the  entropy production rate in non-equilibrium stochastic thermodynamics~\cite{Jarzynski1997-ia, Crooks1999-he}.   Specifically, fitness flux is a measure for the irreversibility of information entropy through adaptation of a population, e.g., for evolution in time-dependent environments. The applicability of fitness flux to co-evolutionary adaptation has been specifically discussed in ref.~\cite{Nourmohammad2016-zg}.  As shown in Appendix~\ref{appendix-ref}, the total rate of entropy production in our bipartite system can be expressed as
\EQA
\nonumber  \Phi &=& \sum_{a'\ge a, v' \ge v}  J_{a' \to a}^{v'\to v } \ln{  \frac{W_{a'\to a}^{v' \to v}\,\, P(a',v' )}{W_{a\to a'}^{v \to v'} \,\,P(a,v)}}\\
\nonumber &=& \underbrace{\sum_{a'\ge a, v }   J_{a' \to a}^{v} \ln{  \frac{W_{a'\to a}^{v }\,\, P(a',v )}{W_{a\to a'}^{v} \,\,P(a,v)}}}_{=\Phi^A}\\
\nonumber && +  \underbrace{\sum_{a, v' \ge v}   J_{a}^{v'\to v }  \ln{  \frac{W_{ a}^{v' \to v}\,\, P(a,v' )}{W_{a}^{v \to v'} \,\,P(a,v)}} }_{=\Phi^V},\\
\label{eq:entropyproduction}
\EEA
where we defined the conditional entropy production rates associated with changes in the antibody subsystem $\Phi^A$ and the viral subsystem $\Phi^V$, conditioned on the state of the other subsystem. The entropy production rate measures the irreversibility of a stochastic process by a Kullback-Leibler divergence between the probability of the forward path (here $W_{a'\to a}^{v' \to v}\,\, p(a',v' )$) and  that of the backward path (here $W_{a\to a'}^{v \to v'} \,\,p(a,v)$).
 Therefore, the entropy production rate $\Phi$, the conditional rates $\Phi_A$ and $\Phi_V$, and by association fitness flux, are non-negative and only zero at equilibrium, where reversibility requires the probabilities of the forward and backward trajectories to be equal. 
 \begin{figure*}[t!]
\includegraphics[width=\textwidth]{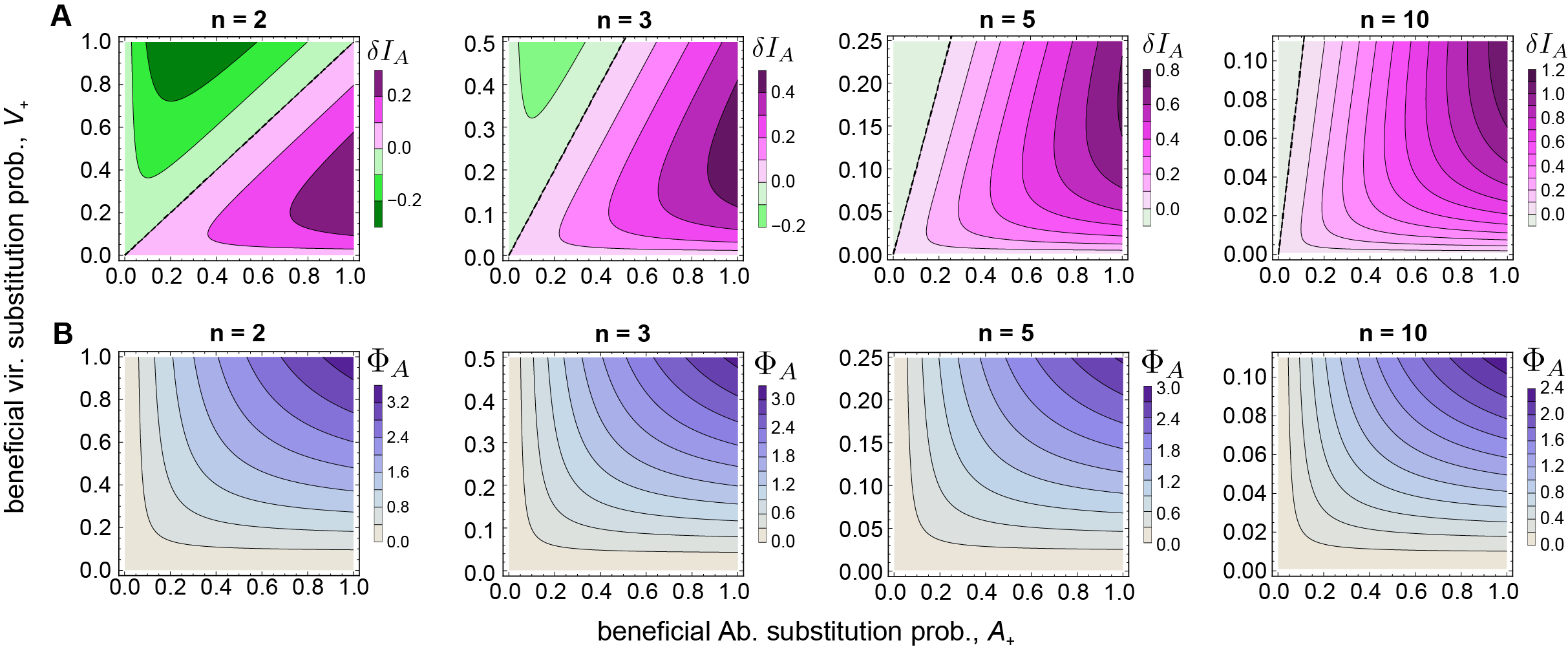}
\caption{{\bf Co-evolutionary information transfer and fitness flux.} Heatmaps show {\bf (A)} the conditional information flux due to antibody evolution $\delta I_A$ (eq.~\eqref{eq:transferflux_bipartite}), which is the measure of the transfer flux from antibodies to viruses~\cite{Nourmohammad2016-zg},  and {\bf (B)} the conditional entropy production rate of antibodies $\Phi_A$ (eq.~\eqref{eq:fitnessflux_bipartite}), which is the measure of antibodies' fitness flux~\cite{Mustonen2010-fj,Nourmohammad2016-zg} in response to viral escape, as a function of the antibody and viral beneficial substitution probabilities, $A_+$, and $V_+$, respectively. The deleterious substitution probabilities are kept fixed $A_- = V_- =  10^{-3}$. Panels show results for different number of states in the balanced system $n_A=n_V=n$. The upper bound for the viral beneficial substitution rate is set using eq.~\ref{eq.boundVplus} to assure that the transition probabilities are appropriately bounded (eq~\eqref{eq:WB_diagonal}).  The dashed lines indicate the diagonal on which $A_+=V_+$.}
\label{Fig:fluxes}
\end{figure*}
In our bipartite model of co-evolution, the conditional entropy production rates quantify the response of each subsystem to the changes caused by the evolution of the other subsystem. Importantly, this response contains the effect of  feedback on the adaptive dynamics of each subsystem during co-evolution. Specifically, the effect of feedback is captured by considering the joint state probability $P(a,v)$, when calculating the entropy production rate in eq.~\eqref{eq:entropyproduction}.  If we were to consider the antibody  subsystem evolving in an independently time-varying viral environment (i.e., by neglecting the feedback), we would have expressed the probability for a transition $a'\to a$ in the environment $v$ by $W_{a'\to a}^{v } P(a')$, where $P(a')$ is the stationary state probability for the system to be in state $a'$, independently of the state of the environment. Similar modification applies to the backward trajectories, and to the transitions in the viral subsystem. Therefore, we can express the contribution from co-evolutionary feedback to the conditional entropy production rate as
\EQA
\nonumber \Phi^A &=& \sum_{a'\ge a, v }   J_{a' \to a}^{v} \ln{  \frac{W_{a'\to a}^{v }\,\, P(v|a' ) P(a')}{W_{a\to a'}^{v} \,\,P(v|a) P(a)}} \\
\nonumber &=& \underbrace{\sum_{a'\ge a, v }   J_{a' \to a}^{v} \ln{  \frac{W_{a'\to a}^{v }\,\, P(a')}{W_{a\to a'}^{v} \,\,P(a)}} }_{\equiv  \Phi^A_0}  -  \underbrace{\sum_{a'\ge a, v }   J_{a' \to a}^{v} \ln{  \frac{ P(v|a ) }{P(v|a') }} }_{ = \delta I^A}, \\
\label{eq:fitnessFluxSplit}
\EEA
where $ \Phi^A_0$ is the entropy production of the antibody subsystem without considering the effect of feedback. Interestingly, the remaining quantity that captures the effect of feedback, $\delta I^A$,  is the conditional mutual information flux~\cite{Horowitz2014-od} due to the change in antibodies, conditioned on the state of viruses (see Appendix~\ref{appendix-ref} for derivation); analogous quantities can be defined for  viruses. The  conditional mutual information flux  $\delta I^A$ quantifies how changes in antibodies impact the phenotypic state and consequently the fitness of viruses; this measure is similar to transfer flux in co-evolving populations, previously introduced in ref.~\cite{Nourmohammad2016-zg}.

In our bipartite co-evolutionary  model, the conditional entropy production rates (i.e., fitness flux) and the conditional mutual information flux associated with each subsystem (i.e., transfer flux) are
\EQA
 \nonumber &&\Phi^A = n(n-1) \frac{\sigma}{Z}  \left[\ln \frac{A_+ }{A_-} +\ln \frac{\pi_\U}{\pi_\BB } \right] \geq 0, \\
 \nonumber &&\Phi^V = n(n-1) \frac{\sigma}{Z}   \left[ \ln \frac{V_+}{V_-} +\ln \frac{\pi_\BB}{\pi_\U } \right]\geq 0, \\
 \label{eq:fitnessflux_bipartite}\\
 \nonumber &&\delta I^A =-\delta I^V = n(n-1)\frac{ \sigma}{Z}  \ln \frac{\pi_\BB}{\pi_\U },  
 \label{eq:transferflux_bipartite}\\
\EEA
where $\sigma = A_+V_+-A_-V_-$, and $Z$ is the normalization constant as in eq.~\eqref{eq:J-Phenotype}. Fig.~\ref{Fig:fluxes} shows the conditional information flux (A) and the entropy production rate  (B) due to  antibody changes for varying levels of antibody and viral substitution probabilities. Notably, the magnitude of both fluxes increases with the number of states, scaling as  $n(n-1)$, which corresponds to the total number of possible transitions. 

As expected, in the steady state, the change in the total mutual information, i.e., $\delta I = \delta I^A+\delta I^V$ is zero. Notably, the sign of the conditional mutual information fluxes, determined by the ratio of the steady-state probabilities $\pi_{\BB}/\pi_{\U}$,  distinguishes between the leader and the follower in this co-evolutionary arms race. When $\pi_{\BB} >\pi_{\U}$, being in a given  bound state is more likely than being in a given unbound state, not withstanding the fact that the number of unbound states is larger. In this case, viral  escape destroys the mutual information $\delta I^V<0$ by  introducing a disorder in an otherwise bound configuration. Antibodies, on the other hand, increase the mutual information $\delta I^A>0$ by learning and aligning themselves to newly evolved viral states, restoring the bound state. In this case, the virus is ``leading'' the evolutionary trajectory while the immune system is ``following."   On the other hand, if $\pi_{\BB} <\pi_{\U}$, then the signs and roles are reversed, and changes in viruses create information, while antibodies erase  information. 

The distinction between the two scenarios leading to the different drivers of the co-evolutionary process is a matter of terminology (i.e., what is called host vs.\ virus or bound vs.\ unbound) rather than reflecting different biologically relevant regimes. As shown in Fig.~\ref{Fig:fluxes}A, the conditional information flux of antibodies $\delta I_A$ is anti-symmetric (i.e., the leader becomes the follower) with respect to the switch in the prominence of the bound vs. unbound state ($\pi_\BB/\pi_\U$); this is captured in Fig.~\ref{Fig:fluxes} by  varying the relative beneficial substitution probabilities $A_+/V_+$, while keeping the deleterious rates constant; see eq.~\eqref{eq.Pstat1}.  As the number of states $n$ increases, the number of possible beneficial states a virus can transition to grows to $(n-1)$, while it remains fixed at one for antibodies (Fig.~\ref{fig:bipartite}B). As a result, the upper bound on the beneficial viral substitution probability $V_+ \leq \frac{1- (n-1)A_-}{n-1}$ (eq.~\ref{eq.boundVplus}) becomes smaller, reducing the size of the region with permutational symmetry, as seen in Fig.~\ref{Fig:fluxes}A.
\begin{figure*}[t!]
\includegraphics[width=\textwidth]{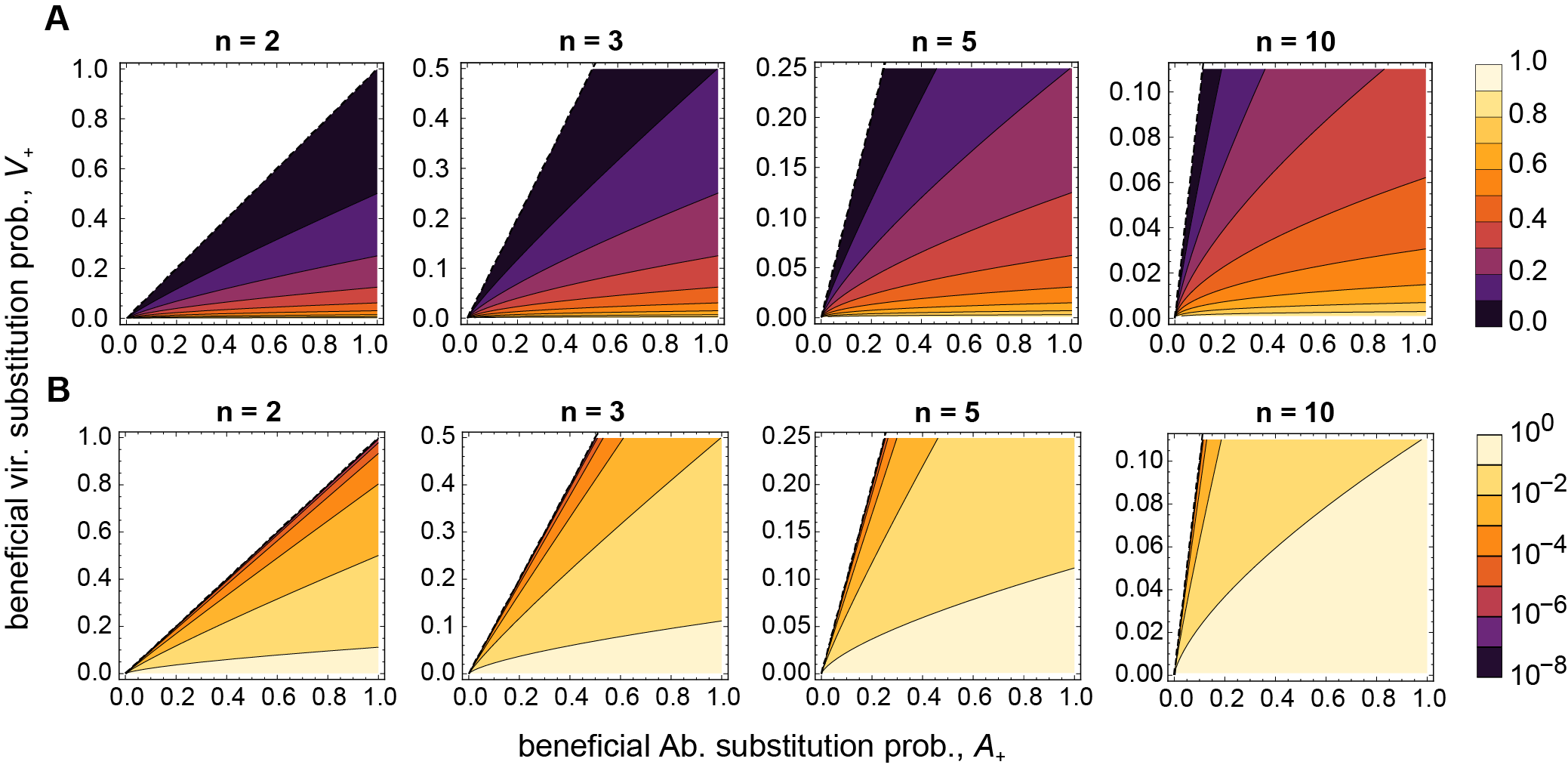}
\caption{{\bf Co-evolutionary efficiency.} {\bf (A)} The efficiency of antibodies $\eta_A$ (eq.~\eqref{eq.EffA}) and {\bf (B)} the total efficiency of the coevolving system $\eta = \eta _A \eta_V$ (eq.~\ref{eq.Eta}) are shown as a function of the antibody and viral beneficial substitution probabilities, $A_+$, and $V_+$, respectively, while keeping the deleterious substitution probabilities fixed $A_- = V_- =  10^{-3}$. Panels show results for different number of states in the balanced system $n_A=n_V=n$. The upper bound for the viral beneficial substitution rate is set using eq.~\ref{eq.boundVplus} to assure that the transition probabilities are appropriately bounded (eq.~\eqref{eq:WB_diagonal}).  The dashed lines indicate the $A_+=V_+$ diagonal. We consider the case where the probability associated with a bound configuration is larger than that of an unbound configuration $\pi_\BB>\pi_\U$, i.e., the region below the dashed lines.
}
\label{Fig:efficiency}
\end{figure*}
 From our biological intuition, the number of possible escape routes for the virus is generally large $n\gg 2$. Moreover, immune repertoires (e.g., antibodies) are diverse enough that for any given pathogenic variant a desired immune solution can be found. Therefore, in the steady state of our co-evolutionary dynamics,  and prior to the escape of the virus, the bound configuration is the more likely state. Thus, we will focus on the case of $\pi_\BB >\pi_\U$ to better interpret our results (below the dashed lines in Fig.~\ref{Fig:fluxes}A). However, similar arguments apply to the opposite case after switching the virus and antibody labels.

\subsection{Efficiency of co-evolutionary information processing}
Let us consider the case in which viruses lead the co-evolutionary arms race by escaping the predominantly bound configurations, and  antibodies follow them. Given that $\delta I^A>0$, eq.~\eqref{eq:fitnessFluxSplit} suggests that the coupling (feedback) between antibodies and viruses leads to a reduction in the rate of adaptation (fitness flux) of antibodies in the co-evolving system compared to that of  a system without feedback, i.e.,  $\Phi^A < \Phi^A_0$. In other words, having  knowledge about the state of the viral ensemble decreases the extent of necessary adaptation (and entropy production) by antibodies. We can define the efficiency of  antibodies  in utilizing the co-evolutionary information by the relative reduction in their entropy production (fitness flux),
\EQ
\eta_A = \frac{\Phi_0^A-\Phi^A}{\Phi^A_0} = \frac{\delta I^A}{\Phi^A+\delta I^A } =  \frac{\ln{\frac{\pi_\BB}{\pi_\U}}}{\ln{\frac{A_+}{A_-}}}  \leq 1.
\label{eq.EffA}
\EE
In the regime where the evolutionary rates of the  antibodies are much larger than that of  viruses $A_+, \,A_- \gg V_+,\,V_-$, we will have $\eta_A \to 1$, implying that the antibodies are very efficient in sensing and adapting to the viral state (Fig.~\ref{Fig:efficiency}A). As  the number of states $n$ increases, the upper bound for the beneficial viral substitution probability becomes smaller (eq.~\ref{eq.boundVplus}), resulting in a broader region for efficient antibody responses (Fig.~\ref{Fig:efficiency}A).

Viruses erase the  information generated by antibodies and adapt to escape the immune system. We can define the efficiency of the virus as the amount of information erasure $|\delta I^V|$ per unit of viral adaptation $ \Phi^V$,
\EQ
\eta_V = \frac{|\delta I^V|}{\Phi^V} = \frac{\ln{\frac{\pi_\BB}{\pi_\U}}} {\ln{\frac{V_+}{V_-}}+\ln{\frac{\pi_\BB}{\pi_\U}}} \leq 1.
\label{eq.EffV}
\EE
 In the regime where the evolutionary rates of  viruses are much larger than that of the antibodies ${V_+, \,V_- \gg A_+,\,A_-}$, we will have $\eta_V \to 1$, implying that  viruses are very efficient in erasing information. 
 
 The efficiency of the joint system can  be defined as 
\EQ 
\eta = \eta_A \eta_V =\frac{1}{\left( 1+\frac{\ln{\frac{V_+}{V_-}}}{\ln{\frac{\pi_\BB}{\pi_\U}}} \right)\frac{\ln{\frac{A_+}{A_-}} }{\ln{\frac{\pi_\BB}{\pi_\U}} } }
\leq 1.
\label{eq.Eta} 
\EE
Note that when the probability of being in an unbound state approaches that of a bound state $\pi_\U \to \pi_\BB$, the efficiency of the joint process goes to its minimum  $\eta\to 0$; see Fig.~\ref{Fig:efficiency}B. In this regime, all the adaptive advantage that antibodies gain are erased by viral escape, driving  the two populations into a continuous evolutionary arms race. Indeed, a balanced adaptive pressure of antibodies and viruses was previously suggested as a criterion to achieve a stable co-evolutionary arms race  in ref.~\cite{Nourmohammad2016-zg}. 

In the regime where $\pi_\BB \gg \pi_\U$ with $A_+/A_- \gg V_+ /V_-$ antibodies can efficiently neutralize viruses, resulting in the maximum efficiency of the joint process that is bounded by  $\eta_\text{max}= \frac{\log A_+/A_-}{\log A_+/A_-+{\log V_+ /V_-}}$. The quantities $A_+/A_-$ and $V_+/V_-$ define the adaptive potential for antibodies and viruses, respectively. In such imbalanced scenario antibody adaptation becomes the dominant process, leading to the elimination of viruses.

It should be noted that the definition of efficiency introduced here is distinct from that of ref.~\cite{Horowitz2014-od}, which considers a physical bipartite system in which the relationship between work, heat dissipation, and (information) entropy production defines the extent of irreversibility in the  system's trajectories. The out-of-equilibrium co-evolutionary process here however does not satisfy a constraint similar to the first law of thermodynamics, and thus physical work is not required for the system to process information, as suggested by the Landauer's principle~\cite{Landauer1961-qn}. The measure of efficiency introduced here is suitable for non-physical stochastic processes, in which the adaptive potential in one system can determine its efficiency in processing the information generated by the other.  
\section{Discussion}
In this study, we introduced a novel framework to model immune-pathogen co-evolution, using a bipartite Markov process to capture the interactions between antibodies and viruses. Our model provides a simplified yet powerful approach to understanding the evolutionary arms race that governs the dynamics of host-pathogen systems. By considering the processes of beneficial, deleterious, and neutral substitutions within this bipartite structure, we established key metrics like information flux, fitness flux, and co-evolutionary efficiency, which enable us to quantify the out-of-equilibrium behavior underlying  immune-pathogen co-evolution, driven by a continuous exchange of information.

A key finding is the identification of co-evolutionary fitness flux and information flux as measures of adaptation and interaction in the coupled immune-pathogen system. The asymmetry in the information flux, measured by the changes in the conditional mutual information due to the evolution of one subpopulation, allows us to distinguish between the leader and follower of a co-evolutionary arms race. In scenarios where viruses rapidly mutate to evade immune detection, they act as the leaders, with antibodies adapting to restore a bound configuration to neutralize them, thus following the viral evolution. 

Our model highlights the use of co-evolutionary efficiency in determining  how each population exploits the information to adapt more effectively towards their desired states. Specifically, we define efficiency as the relative reduction in the information entropy production of each population due to the feedback from the other. Antibodies, for example, benefit from this feedback, reducing their rate of adaptation when they have information about the state of the viral population. 

Prior work has  focused on how viral mutation rates are set by the balance between the benefits of increased mutation for immune evasion and the costs of harmful off-target mutations, known as genetic load~\cite{Bull2007-nj,Koelle2015-kg,Swanstrom2022-gk}. Our framework suggests an alternative approach to understanding how mutations rates, or more precisely substitution rates, are to be modulated. We show that a co-evolutionary system’s overall efficiency is determined by the balance between the adaptation rates of the virus and the antibody populations. Specifically, when the relative beneficial to deleterious substitution rates for both populations are comparable the joint system reaches it minimum efficiency. In this regime, the adaptive advantage that antibodies gain is erased by viral escape, driving the two populations to maintain a constant evolutionary arms race.

While it is challenging to directly measure the co-evolutionary efficiency of viral-immune systems, insights can be gained through the study of endogenous viral elements (EVEs). EVEs are segments of viral DNA that have become integrated into the genomes of their hosts and are transmitted across generations,  serving as fossil records of ancient viruses over millions of years. Once integrated, EVEs remain highly stable and can be used to trace the evolutionary origins of different viruses through phylogenetic analyses of host populations. For instance, although HIV-1 was recently introduced into the human population, its origins can be traced back to endogenous lentiviruses in rabbits and other mammals, with its close relative, the simian immunodeficiency virus, dating back over 10,000 years.

Analysis of  sequence divergence of mammalian viruses reveals that over short timescales viral evolution tends to exhibit rapid rates of sequence divergence. In contrast, when divergence is measured from EVEs over much longer timescales, the observed evolutionary rates are significantly slower, approaching that of the host's genome~\cite{Simmonds2019-su}. The inherent biases in these data-- e.g., sequences  that can be aligned over long evolutionary times ought to be similar-- make it difficult to draw definitive conclusions about the evolution of viral substitution rates. Nonetheless, this pattern raises the question as whether viruses that have established long-term coexistence with their hosts have evolved to balance  their beneficial and deleterious substitution rates with that of the host's immune updates.  Our measure of co-evolutionary efficiency may guide studies of EVEs in host genomes, especially from ancient DNA, to address these questions.

Prior work on stochastic thermodynamics has extended the concept of efficiency for information exchange~\cite{Horowitz2014-od,Leighton2023-zi}. However, they  consider physical systems, in which the  energy  is conserved (first law of thermodynamics), and therefore, quantities like work and heat are well-defined. Immune-pathogen coevolution is a specific example of an interacting non-physical stochastic system with information exchange that does not follow the conservation law for an energy-like quantity. Therefore, the notion of efficiency introduced here can be used more broadly for non-physical stochastic systems like those in economics, financial markets, linguistics, and artificial neural networks. 

Our model presents a highly simplified description of immune-pathogen co-evolution. Combining this framework with a more realistic description of the evolution and ecology of interacting populations could shed light on how factors such as population diversity, demographic structures, immune memory and cross-immunity  influence the co-evolutionary process. Additionally, while we focused on the symmetric case where the number of antibody and viral states are balanced, the imbalanced case could be explored further to understand evolutionary distraction for antibodies.

\section{Acknowledgment}
This work has been supported by the CAREER award from the National Science Foundation (grant No: 2045054), and the MPRG funding from the Max Planck Society to A.N.. This work is also supported, in part, through the Department of Physics and the College of Arts and Sciences at the University of Washington.


\begin{thebibliography}{43}%
\makeatletter
\providecommand \@ifxundefined [1]{%
 \@ifx{#1\undefined}
}%
\providecommand \@ifnum [1]{%
 \ifnum #1\expandafter \@firstoftwo
 \else \expandafter \@secondoftwo
 \fi
}%
\providecommand \@ifx [1]{%
 \ifx #1\expandafter \@firstoftwo
 \else \expandafter \@secondoftwo
 \fi
}%
\providecommand \natexlab [1]{#1}%
\providecommand \enquote  [1]{``#1''}%
\providecommand \bibnamefont  [1]{#1}%
\providecommand \bibfnamefont [1]{#1}%
\providecommand \citenamefont [1]{#1}%
\providecommand \href@noop [0]{\@secondoftwo}%
\providecommand \href [0]{\begingroup \@sanitize@url \@href}%
\providecommand \@href[1]{\@@startlink{#1}\@@href}%
\providecommand \@@href[1]{\endgroup#1\@@endlink}%
\providecommand \@sanitize@url [0]{\catcode `\\12\catcode `\$12\catcode
  `\&12\catcode `\#12\catcode `\^12\catcode `\_12\catcode `\%12\relax}%
\providecommand \@@startlink[1]{}%
\providecommand \@@endlink[0]{}%
\providecommand \url  [0]{\begingroup\@sanitize@url \@url }%
\providecommand \@url [1]{\endgroup\@href {#1}{\urlprefix }}%
\providecommand \urlprefix  [0]{URL }%
\providecommand \Eprint [0]{\href }%
\providecommand \doibase [0]{http://dx.doi.org/}%
\providecommand \selectlanguage [0]{\@gobble}%
\providecommand \bibinfo  [0]{\@secondoftwo}%
\providecommand \bibfield  [0]{\@secondoftwo}%
\providecommand \translation [1]{[#1]}%
\providecommand \BibitemOpen [0]{}%
\providecommand \bibitemStop [0]{}%
\providecommand \bibitemNoStop [0]{.\EOS\space}%
\providecommand \EOS [0]{\spacefactor3000\relax}%
\providecommand \BibitemShut  [1]{\csname bibitem#1\endcsname}%
\let\auto@bib@innerbib\@empty
\bibitem [{\citenamefont {Janeway}\ \emph {et~al.}(2001)\citenamefont
  {Janeway}, \citenamefont {Travers}, \citenamefont {Walport},\ and\
  \citenamefont {Shlomchik}}]{Janeway2001-ws}%
  \BibitemOpen
  \bibfield  {author} {\bibinfo {author} {\bibfnamefont {C.~A.}\ \bibnamefont
  {Janeway}, \bibfnamefont {Jr}}, \bibinfo {author} {\bibfnamefont
  {P.}~\bibnamefont {Travers}}, \bibinfo {author} {\bibfnamefont
  {M.}~\bibnamefont {Walport}}, \ and\ \bibinfo {author} {\bibfnamefont
  {M.~J.}\ \bibnamefont {Shlomchik}},\ }\href@noop {} {\emph {\bibinfo {title}
  {Immunobiology}}}\ (\bibinfo  {publisher} {Garland Science},\ \bibinfo {year}
  {2001})\BibitemShut {NoStop}%
\bibitem [{\citenamefont {Hampton}\ \emph {et~al.}(2020)\citenamefont
  {Hampton}, \citenamefont {Watson},\ and\ \citenamefont
  {Fineran}}]{Hampton2020-ds}%
  \BibitemOpen
  \bibfield  {author} {\bibinfo {author} {\bibfnamefont {H.~G.}\ \bibnamefont
  {Hampton}}, \bibinfo {author} {\bibfnamefont {B.~N.~J.}\ \bibnamefont
  {Watson}}, \ and\ \bibinfo {author} {\bibfnamefont {P.~C.}\ \bibnamefont
  {Fineran}},\ }\href {\doibase 10.1038/s41586-019-1894-8} {\bibfield
  {journal} {\bibinfo  {journal} {Nature}\ }\textbf {\bibinfo {volume} {577}},\
  \bibinfo {pages} {327} (\bibinfo {year} {2020})}\BibitemShut {NoStop}%
\bibitem [{\citenamefont {Bradde}\ \emph {et~al.}(2020)\citenamefont {Bradde},
  \citenamefont {Nourmohammad}, \citenamefont {Goyal},\ and\ \citenamefont
  {Balasubramanian}}]{Bradde2020-iy}%
  \BibitemOpen
  \bibfield  {author} {\bibinfo {author} {\bibfnamefont {S.}~\bibnamefont
  {Bradde}}, \bibinfo {author} {\bibfnamefont {A.}~\bibnamefont
  {Nourmohammad}}, \bibinfo {author} {\bibfnamefont {S.}~\bibnamefont {Goyal}},
  \ and\ \bibinfo {author} {\bibfnamefont {V.}~\bibnamefont
  {Balasubramanian}},\ }\href {\doibase 10.1073/pnas.1903666117} {\bibfield
  {journal} {\bibinfo  {journal} {Proc. Natl. Acad. Sci. U. S. A.}\ }\textbf
  {\bibinfo {volume} {117}},\ \bibinfo {pages} {5144} (\bibinfo {year}
  {2020})}\BibitemShut {NoStop}%
\bibitem [{\citenamefont {Watson}\ \emph {et~al.}(2021)\citenamefont {Watson},
  \citenamefont {Steens}, \citenamefont {Staals}, \citenamefont {Westra},\ and\
  \citenamefont {van Houte}}]{Watson2021-ul}%
  \BibitemOpen
  \bibfield  {author} {\bibinfo {author} {\bibfnamefont {B.~N.~J.}\
  \bibnamefont {Watson}}, \bibinfo {author} {\bibfnamefont {J.~A.}\
  \bibnamefont {Steens}}, \bibinfo {author} {\bibfnamefont {R.~H.~J.}\
  \bibnamefont {Staals}}, \bibinfo {author} {\bibfnamefont {E.~R.}\
  \bibnamefont {Westra}}, \ and\ \bibinfo {author} {\bibfnamefont
  {S.}~\bibnamefont {van Houte}},\ }\href {\doibase 10.1016/j.chom.2021.03.018}
  {\bibfield  {journal} {\bibinfo  {journal} {Cell Host Microbe}\ }\textbf
  {\bibinfo {volume} {29}},\ \bibinfo {pages} {715} (\bibinfo {year}
  {2021})}\BibitemShut {NoStop}%
\bibitem [{\citenamefont {Nourmohammad}\ \emph {et~al.}(2016)\citenamefont
  {Nourmohammad}, \citenamefont {Otwinowski},\ and\ \citenamefont
  {Plotkin}}]{Nourmohammad2016-zg}%
  \BibitemOpen
  \bibfield  {author} {\bibinfo {author} {\bibfnamefont {A.}~\bibnamefont
  {Nourmohammad}}, \bibinfo {author} {\bibfnamefont {J.}~\bibnamefont
  {Otwinowski}}, \ and\ \bibinfo {author} {\bibfnamefont {J.~B.}\ \bibnamefont
  {Plotkin}},\ }\href {\doibase 10.1371/journal.pgen.1006171} {\bibfield
  {journal} {\bibinfo  {journal} {PLoS Genet.}\ }\textbf {\bibinfo {volume}
  {12}},\ \bibinfo {pages} {e1006171} (\bibinfo {year} {2016})}\BibitemShut
  {NoStop}%
\bibitem [{\citenamefont {Liao}\ \emph {et~al.}(2013)\citenamefont {Liao},
  \citenamefont {Lynch}, \citenamefont {Zhou}, \citenamefont {Gao},
  \citenamefont {Alam}, \citenamefont {Boyd}, \citenamefont {Fire},
  \citenamefont {Roskin}, \citenamefont {Schramm}, \citenamefont {Zhang},
  \citenamefont {Zhu}, \citenamefont {Shapiro}, \citenamefont {Mullikin},
  \citenamefont {Gnanakaran}, \citenamefont {Hraber}, \citenamefont {Wiehe},
  \citenamefont {Kelsoe}, \citenamefont {Yang}, \citenamefont {Xia},
  \citenamefont {Montefiori}, \citenamefont {Parks}, \citenamefont {Lloyd},
  \citenamefont {Scearce}, \citenamefont {Soderberg}, \citenamefont {Cohen},
  \citenamefont {Kamanga}, \citenamefont {Louder}, \citenamefont {Tran},
  \citenamefont {Chen}, \citenamefont {Cai}, \citenamefont {Chen},
  \citenamefont {Moquin}, \citenamefont {Du}, \citenamefont {Joyce},
  \citenamefont {Srivatsan}, \citenamefont {Zhang}, \citenamefont {Zheng},
  \citenamefont {Shaw}, \citenamefont {Hahn}, \citenamefont {Kepler},
  \citenamefont {Korber}, \citenamefont {Kwong}, \citenamefont {Mascola},\ and\
  \citenamefont {Haynes}}]{Liao2013-ga}%
  \BibitemOpen
  \bibfield  {author} {\bibinfo {author} {\bibfnamefont {H.-X.}\ \bibnamefont
  {Liao}}, \bibinfo {author} {\bibfnamefont {R.}~\bibnamefont {Lynch}},
  \bibinfo {author} {\bibfnamefont {T.}~\bibnamefont {Zhou}}, \bibinfo {author}
  {\bibfnamefont {F.}~\bibnamefont {Gao}}, \bibinfo {author} {\bibfnamefont
  {S.~M.}\ \bibnamefont {Alam}}, \bibinfo {author} {\bibfnamefont {S.~D.}\
  \bibnamefont {Boyd}}, \bibinfo {author} {\bibfnamefont {A.~Z.}\ \bibnamefont
  {Fire}}, \bibinfo {author} {\bibfnamefont {K.~M.}\ \bibnamefont {Roskin}},
  \bibinfo {author} {\bibfnamefont {C.~A.}\ \bibnamefont {Schramm}}, \bibinfo
  {author} {\bibfnamefont {Z.}~\bibnamefont {Zhang}}, \bibinfo {author}
  {\bibfnamefont {J.}~\bibnamefont {Zhu}}, \bibinfo {author} {\bibfnamefont
  {L.}~\bibnamefont {Shapiro}}, \bibinfo {author} {\bibfnamefont {J.~C.}\
  \bibnamefont {Mullikin}}, \bibinfo {author} {\bibfnamefont {S.}~\bibnamefont
  {Gnanakaran}}, \bibinfo {author} {\bibfnamefont {P.}~\bibnamefont {Hraber}},
  \bibinfo {author} {\bibfnamefont {K.}~\bibnamefont {Wiehe}}, \bibinfo
  {author} {\bibfnamefont {G.}~\bibnamefont {Kelsoe}}, \bibinfo {author}
  {\bibfnamefont {G.}~\bibnamefont {Yang}}, \bibinfo {author} {\bibfnamefont
  {S.-M.}\ \bibnamefont {Xia}}, \bibinfo {author} {\bibfnamefont {D.~C.}\
  \bibnamefont {Montefiori}}, \bibinfo {author} {\bibfnamefont
  {R.}~\bibnamefont {Parks}}, \bibinfo {author} {\bibfnamefont {K.~E.}\
  \bibnamefont {Lloyd}}, \bibinfo {author} {\bibfnamefont {R.~M.}\ \bibnamefont
  {Scearce}}, \bibinfo {author} {\bibfnamefont {K.~A.}\ \bibnamefont
  {Soderberg}}, \bibinfo {author} {\bibfnamefont {M.}~\bibnamefont {Cohen}},
  \bibinfo {author} {\bibfnamefont {G.}~\bibnamefont {Kamanga}}, \bibinfo
  {author} {\bibfnamefont {M.~K.}\ \bibnamefont {Louder}}, \bibinfo {author}
  {\bibfnamefont {L.~M.}\ \bibnamefont {Tran}}, \bibinfo {author}
  {\bibfnamefont {Y.}~\bibnamefont {Chen}}, \bibinfo {author} {\bibfnamefont
  {F.}~\bibnamefont {Cai}}, \bibinfo {author} {\bibfnamefont {S.}~\bibnamefont
  {Chen}}, \bibinfo {author} {\bibfnamefont {S.}~\bibnamefont {Moquin}},
  \bibinfo {author} {\bibfnamefont {X.}~\bibnamefont {Du}}, \bibinfo {author}
  {\bibfnamefont {M.~G.}\ \bibnamefont {Joyce}}, \bibinfo {author}
  {\bibfnamefont {S.}~\bibnamefont {Srivatsan}}, \bibinfo {author}
  {\bibfnamefont {B.}~\bibnamefont {Zhang}}, \bibinfo {author} {\bibfnamefont
  {A.}~\bibnamefont {Zheng}}, \bibinfo {author} {\bibfnamefont {G.~M.}\
  \bibnamefont {Shaw}}, \bibinfo {author} {\bibfnamefont {B.~H.}\ \bibnamefont
  {Hahn}}, \bibinfo {author} {\bibfnamefont {T.~B.}\ \bibnamefont {Kepler}},
  \bibinfo {author} {\bibfnamefont {B.~T.~M.}\ \bibnamefont {Korber}}, \bibinfo
  {author} {\bibfnamefont {P.~D.}\ \bibnamefont {Kwong}}, \bibinfo {author}
  {\bibfnamefont {J.~R.}\ \bibnamefont {Mascola}}, \ and\ \bibinfo {author}
  {\bibfnamefont {B.~F.}\ \bibnamefont {Haynes}},\ }\href {\doibase
  10.1038/nature12053} {\bibfield  {journal} {\bibinfo  {journal} {Nature}\
  }\textbf {\bibinfo {volume} {496}},\ \bibinfo {pages} {469} (\bibinfo {year}
  {2013})}\BibitemShut {NoStop}%
\bibitem [{\citenamefont {Luo}\ and\ \citenamefont
  {Perelson}(2015)}]{Luo2015-mj}%
  \BibitemOpen
  \bibfield  {author} {\bibinfo {author} {\bibfnamefont {S.}~\bibnamefont
  {Luo}}\ and\ \bibinfo {author} {\bibfnamefont {A.~S.}\ \bibnamefont
  {Perelson}},\ }\href {\doibase 10.1073/pnas.1505207112} {\bibfield  {journal}
  {\bibinfo  {journal} {Proceedings of the National Academy of Sciences}\
  }\textbf {\bibinfo {volume} {112}},\ \bibinfo {pages} {11654} (\bibinfo
  {year} {2015})},\ \Eprint
  {http://arxiv.org/abs/https://www.pnas.org/doi/pdf/10.1073/pnas.1505207112}
  {https://www.pnas.org/doi/pdf/10.1073/pnas.1505207112} \BibitemShut {NoStop}%
\bibitem [{\citenamefont {Wang}\ \emph {et~al.}(2015)\citenamefont {Wang},
  \citenamefont {Mata-Fink}, \citenamefont {Kriegsman}, \citenamefont {Hanson},
  \citenamefont {Irvine}, \citenamefont {Eisen}, \citenamefont {Burton},
  \citenamefont {Wittrup}, \citenamefont {Kardar},\ and\ \citenamefont
  {Chakraborty}}]{Wang2015-rf}%
  \BibitemOpen
  \bibfield  {author} {\bibinfo {author} {\bibfnamefont {S.}~\bibnamefont
  {Wang}}, \bibinfo {author} {\bibfnamefont {J.}~\bibnamefont {Mata-Fink}},
  \bibinfo {author} {\bibfnamefont {B.}~\bibnamefont {Kriegsman}}, \bibinfo
  {author} {\bibfnamefont {M.}~\bibnamefont {Hanson}}, \bibinfo {author}
  {\bibfnamefont {D.~J.}\ \bibnamefont {Irvine}}, \bibinfo {author}
  {\bibfnamefont {H.~N.}\ \bibnamefont {Eisen}}, \bibinfo {author}
  {\bibfnamefont {D.~R.}\ \bibnamefont {Burton}}, \bibinfo {author}
  {\bibfnamefont {K.~D.}\ \bibnamefont {Wittrup}}, \bibinfo {author}
  {\bibfnamefont {M.}~\bibnamefont {Kardar}}, \ and\ \bibinfo {author}
  {\bibfnamefont {A.~K.}\ \bibnamefont {Chakraborty}},\ }\href {\doibase
  10.1016/j.cell.2015.01.027} {\bibfield  {journal} {\bibinfo  {journal}
  {Cell}\ }\textbf {\bibinfo {volume} {160}},\ \bibinfo {pages} {785} (\bibinfo
  {year} {2015})}\BibitemShut {NoStop}%
\bibitem [{\citenamefont {Nourmohammad}\ \emph {et~al.}(2019)\citenamefont
  {Nourmohammad}, \citenamefont {Otwinowski}, \citenamefont {Łuksza},
  \citenamefont {Mora},\ and\ \citenamefont {Walczak}}]{Nourmohammad2019-lk}%
  \BibitemOpen
  \bibfield  {author} {\bibinfo {author} {\bibfnamefont {A.}~\bibnamefont
  {Nourmohammad}}, \bibinfo {author} {\bibfnamefont {J.}~\bibnamefont
  {Otwinowski}}, \bibinfo {author} {\bibfnamefont {M.}~\bibnamefont {Łuksza}},
  \bibinfo {author} {\bibfnamefont {T.}~\bibnamefont {Mora}}, \ and\ \bibinfo
  {author} {\bibfnamefont {A.~M.}\ \bibnamefont {Walczak}},\ }\href {\doibase
  10.1093/molbev/msz143} {\bibfield  {journal} {\bibinfo  {journal} {Mol. Biol.
  Evol.}\ }\textbf {\bibinfo {volume} {36}},\ \bibinfo {pages} {2184} (\bibinfo
  {year} {2019})}\BibitemShut {NoStop}%
\bibitem [{\citenamefont {Strauli}\ \emph {et~al.}(2019)\citenamefont
  {Strauli}, \citenamefont {Fryer}, \citenamefont {Pham}, \citenamefont
  {Abdel-Mohsen}, \citenamefont {Facente}, \citenamefont {Pilcher},
  \citenamefont {Pennings}, \citenamefont {Pillai},\ and\ \citenamefont
  {Hernandez}}]{Strauli2019-xx}%
  \BibitemOpen
  \bibfield  {author} {\bibinfo {author} {\bibfnamefont {N.}~\bibnamefont
  {Strauli}}, \bibinfo {author} {\bibfnamefont {E.~K.}\ \bibnamefont {Fryer}},
  \bibinfo {author} {\bibfnamefont {O.}~\bibnamefont {Pham}}, \bibinfo {author}
  {\bibfnamefont {M.}~\bibnamefont {Abdel-Mohsen}}, \bibinfo {author}
  {\bibfnamefont {S.~N.}\ \bibnamefont {Facente}}, \bibinfo {author}
  {\bibfnamefont {C.}~\bibnamefont {Pilcher}}, \bibinfo {author} {\bibfnamefont
  {P.}~\bibnamefont {Pennings}}, \bibinfo {author} {\bibfnamefont
  {S.}~\bibnamefont {Pillai}}, \ and\ \bibinfo {author} {\bibfnamefont {R.~D.}\
  \bibnamefont {Hernandez}},\ }\href {\doibase 10.1101/646968} {\bibfield
  {journal} {\bibinfo  {journal} {bioRxiv}\ ,\ \bibinfo {pages} {646968}}
  (\bibinfo {year} {2019})}\BibitemShut {NoStop}%
\bibitem [{\citenamefont {Mazzolini}\ \emph {et~al.}(2023)\citenamefont
  {Mazzolini}, \citenamefont {Mora},\ and\ \citenamefont
  {Walczak}}]{Mazzolini2023-sd}%
  \BibitemOpen
  \bibfield  {author} {\bibinfo {author} {\bibfnamefont {A.}~\bibnamefont
  {Mazzolini}}, \bibinfo {author} {\bibfnamefont {T.}~\bibnamefont {Mora}}, \
  and\ \bibinfo {author} {\bibfnamefont {A.~M.}\ \bibnamefont {Walczak}},\
  }\href {\doibase 10.1098/rstb.2022.0056} {\bibfield  {journal} {\bibinfo
  {journal} {Philos. Trans. R. Soc. Lond. B Biol. Sci.}\ }\textbf {\bibinfo
  {volume} {378}},\ \bibinfo {pages} {20220056} (\bibinfo {year}
  {2023})}\BibitemShut {NoStop}%
\bibitem [{\citenamefont {Victora}\ and\ \citenamefont
  {Nussenzweig}(2022)}]{Victora2022-an}%
  \BibitemOpen
  \bibfield  {author} {\bibinfo {author} {\bibfnamefont {G.~D.}\ \bibnamefont
  {Victora}}\ and\ \bibinfo {author} {\bibfnamefont {M.~C.}\ \bibnamefont
  {Nussenzweig}},\ }\href {\doibase 10.1146/annurev-immunol-120419-022408}
  {\bibfield  {journal} {\bibinfo  {journal} {Annu. Rev. Immunol.}\ }\textbf
  {\bibinfo {volume} {40}},\ \bibinfo {pages} {413} (\bibinfo {year}
  {2022})}\BibitemShut {NoStop}%
\bibitem [{\citenamefont {Barrat-Charlaix}\ and\ \citenamefont
  {Neher}(2024)}]{Barrat-Charlaix2024-gx}%
  \BibitemOpen
  \bibfield  {author} {\bibinfo {author} {\bibfnamefont {P.}~\bibnamefont
  {Barrat-Charlaix}}\ and\ \bibinfo {author} {\bibfnamefont {R.~A.}\
  \bibnamefont {Neher}},\ }\href {\doibase 10.7554/elife.97350.1} {\bibfield
  {journal} {\bibinfo  {journal} {eLife}\ } (\bibinfo {year} {2024}),\
  10.7554/elife.97350.1}\BibitemShut {NoStop}%
\bibitem [{\citenamefont {Meijers}\ \emph {et~al.}(2023)\citenamefont
  {Meijers}, \citenamefont {Ruchnewitz}, \citenamefont {Eberhardt},
  \citenamefont {Łuksza},\ and\ \citenamefont {Lässig}}]{Meijers2023-sr}%
  \BibitemOpen
  \bibfield  {author} {\bibinfo {author} {\bibfnamefont {M.}~\bibnamefont
  {Meijers}}, \bibinfo {author} {\bibfnamefont {D.}~\bibnamefont {Ruchnewitz}},
  \bibinfo {author} {\bibfnamefont {J.}~\bibnamefont {Eberhardt}}, \bibinfo
  {author} {\bibfnamefont {M.}~\bibnamefont {Łuksza}}, \ and\ \bibinfo
  {author} {\bibfnamefont {M.}~\bibnamefont {Lässig}},\ }\href {\doibase
  10.1016/j.cell.2023.09.022} {\bibfield  {journal} {\bibinfo  {journal}
  {Cell}\ }\textbf {\bibinfo {volume} {186}},\ \bibinfo {pages} {5151}
  (\bibinfo {year} {2023})}\BibitemShut {NoStop}%
\bibitem [{\citenamefont {Marchi}\ \emph {et~al.}(2021)\citenamefont {Marchi},
  \citenamefont {Lässig}, \citenamefont {Walczak},\ and\ \citenamefont
  {Mora}}]{Marchi2021-xd}%
  \BibitemOpen
  \bibfield  {author} {\bibinfo {author} {\bibfnamefont {J.}~\bibnamefont
  {Marchi}}, \bibinfo {author} {\bibfnamefont {M.}~\bibnamefont {Lässig}},
  \bibinfo {author} {\bibfnamefont {A.~M.}\ \bibnamefont {Walczak}}, \ and\
  \bibinfo {author} {\bibfnamefont {T.}~\bibnamefont {Mora}},\ }\href {\doibase
  10.1073/pnas.2103398118} {\bibfield  {journal} {\bibinfo  {journal} {Proc.
  Natl. Acad. Sci. U. S. A.}\ }\textbf {\bibinfo {volume} {118}} (\bibinfo
  {year} {2021}),\ 10.1073/pnas.2103398118}\BibitemShut {NoStop}%
\bibitem [{\citenamefont {Schnaack}\ and\ \citenamefont
  {Nourmohammad}(2021)}]{Schnaack2021-fo}%
  \BibitemOpen
  \bibfield  {author} {\bibinfo {author} {\bibfnamefont {O.~H.}\ \bibnamefont
  {Schnaack}}\ and\ \bibinfo {author} {\bibfnamefont {A.}~\bibnamefont
  {Nourmohammad}},\ }\href {\doibase 10.7554/eLife.61346} {\bibfield  {journal}
  {\bibinfo  {journal} {Elife}\ }\textbf {\bibinfo {volume} {10}} (\bibinfo
  {year} {2021}),\ 10.7554/eLife.61346}\BibitemShut {NoStop}%
\bibitem [{\citenamefont {Schnaack}\ \emph {et~al.}(2022)\citenamefont
  {Schnaack}, \citenamefont {Peliti},\ and\ \citenamefont
  {Nourmohammad}}]{Schnaack2022-if}%
  \BibitemOpen
  \bibfield  {author} {\bibinfo {author} {\bibfnamefont {O.~H.}\ \bibnamefont
  {Schnaack}}, \bibinfo {author} {\bibfnamefont {L.}~\bibnamefont {Peliti}}, \
  and\ \bibinfo {author} {\bibfnamefont {A.}~\bibnamefont {Nourmohammad}},\
  }\href {\doibase 10.1103/PhysRevX.12.021063} {\bibfield  {journal} {\bibinfo
  {journal} {Phys. Rev. X}\ }\textbf {\bibinfo {volume} {12}},\ \bibinfo
  {pages} {021063} (\bibinfo {year} {2022})}\BibitemShut {NoStop}%
\bibitem [{\citenamefont {Chardès}\ \emph {et~al.}(2023)\citenamefont
  {Chardès}, \citenamefont {Mazzolini}, \citenamefont {Mora},\ and\
  \citenamefont {Walczak}}]{Chardes2023-if}%
  \BibitemOpen
  \bibfield  {author} {\bibinfo {author} {\bibfnamefont {V.}~\bibnamefont
  {Chardès}}, \bibinfo {author} {\bibfnamefont {A.}~\bibnamefont {Mazzolini}},
  \bibinfo {author} {\bibfnamefont {T.}~\bibnamefont {Mora}}, \ and\ \bibinfo
  {author} {\bibfnamefont {A.~M.}\ \bibnamefont {Walczak}},\ }\href {\doibase
  10.1073/pnas.2307712120} {\bibfield  {journal} {\bibinfo  {journal} {Proc.
  Natl. Acad. Sci. U. S. A.}\ }\textbf {\bibinfo {volume} {120}},\ \bibinfo
  {pages} {e2307712120} (\bibinfo {year} {2023})}\BibitemShut {NoStop}%
\bibitem [{\citenamefont {Barton}\ and\ \citenamefont
  {de~Vladar}(2009)}]{Barton2009-ak}%
  \BibitemOpen
  \bibfield  {author} {\bibinfo {author} {\bibfnamefont {N.~H.}\ \bibnamefont
  {Barton}}\ and\ \bibinfo {author} {\bibfnamefont {H.~P.}\ \bibnamefont
  {de~Vladar}},\ }\href {\doibase 10.1534/genetics.108.099309} {\bibfield
  {journal} {\bibinfo  {journal} {Genetics}\ }\textbf {\bibinfo {volume}
  {181}},\ \bibinfo {pages} {997} (\bibinfo {year} {2009})}\BibitemShut
  {NoStop}%
\bibitem [{\citenamefont {Nourmohammad}\ \emph
  {et~al.}(2013{\natexlab{a}})\citenamefont {Nourmohammad}, \citenamefont
  {Schiffels},\ and\ \citenamefont {Lässig}}]{Nourmohammad2013-hs}%
  \BibitemOpen
  \bibfield  {author} {\bibinfo {author} {\bibfnamefont {A.}~\bibnamefont
  {Nourmohammad}}, \bibinfo {author} {\bibfnamefont {S.}~\bibnamefont
  {Schiffels}}, \ and\ \bibinfo {author} {\bibfnamefont {M.}~\bibnamefont
  {Lässig}},\ }\href {\doibase 10.1088/1742-5468/2013/01/P01012} {\bibfield
  {journal} {\bibinfo  {journal} {J. Stat. Mech.}\ }\textbf {\bibinfo {volume}
  {2013}},\ \bibinfo {pages} {P01012} (\bibinfo {year}
  {2013}{\natexlab{a}})}\BibitemShut {NoStop}%
\bibitem [{\citenamefont {Nourmohammad}\ \emph
  {et~al.}(2013{\natexlab{b}})\citenamefont {Nourmohammad}, \citenamefont
  {Held},\ and\ \citenamefont {Lässig}}]{Nourmohammad2013-tm}%
  \BibitemOpen
  \bibfield  {author} {\bibinfo {author} {\bibfnamefont {A.}~\bibnamefont
  {Nourmohammad}}, \bibinfo {author} {\bibfnamefont {T.}~\bibnamefont {Held}},
  \ and\ \bibinfo {author} {\bibfnamefont {M.}~\bibnamefont {Lässig}},\ }\href
  {\doibase 10.1016/j.gde.2013.11.001} {\bibfield  {journal} {\bibinfo
  {journal} {Curr. Opin. Genet. Dev.}\ }\textbf {\bibinfo {volume} {23}},\
  \bibinfo {pages} {684} (\bibinfo {year} {2013}{\natexlab{b}})}\BibitemShut
  {NoStop}%
\bibitem [{\citenamefont {Sella}\ and\ \citenamefont
  {Hirsh}(2005)}]{Sella2005-br}%
  \BibitemOpen
  \bibfield  {author} {\bibinfo {author} {\bibfnamefont {G.}~\bibnamefont
  {Sella}}\ and\ \bibinfo {author} {\bibfnamefont {A.~E.}\ \bibnamefont
  {Hirsh}},\ }\href {\doibase 10.1073/pnas.0501865102} {\bibfield  {journal}
  {\bibinfo  {journal} {Proc. Natl. Acad. Sci. U. S. A.}\ }\textbf {\bibinfo
  {volume} {102}},\ \bibinfo {pages} {9541} (\bibinfo {year}
  {2005})}\BibitemShut {NoStop}%
\bibitem [{\citenamefont {Mustonen}\ and\ \citenamefont
  {Lässig}(2005)}]{Mustonen2005-yb}%
  \BibitemOpen
  \bibfield  {author} {\bibinfo {author} {\bibfnamefont {V.}~\bibnamefont
  {Mustonen}}\ and\ \bibinfo {author} {\bibfnamefont {M.}~\bibnamefont
  {Lässig}},\ }\href {\doibase 10.1073/pnas.0505537102} {\bibfield  {journal}
  {\bibinfo  {journal} {Proc. Natl. Acad. Sci. U. S. A.}\ }\textbf {\bibinfo
  {volume} {102}},\ \bibinfo {pages} {15936} (\bibinfo {year}
  {2005})}\BibitemShut {NoStop}%
\bibitem [{\citenamefont {Mustonen}\ and\ \citenamefont
  {Lässig}(2009)}]{Mustonen2009-uh}%
  \BibitemOpen
  \bibfield  {author} {\bibinfo {author} {\bibfnamefont {V.}~\bibnamefont
  {Mustonen}}\ and\ \bibinfo {author} {\bibfnamefont {M.}~\bibnamefont
  {Lässig}},\ }\href {\doibase 10.1016/j.tig.2009.01.002} {\bibfield
  {journal} {\bibinfo  {journal} {Trends Genet.}\ }\textbf {\bibinfo {volume}
  {25}},\ \bibinfo {pages} {111} (\bibinfo {year} {2009})}\BibitemShut
  {NoStop}%
\bibitem [{\citenamefont {Mustonen}\ and\ \citenamefont
  {Lässig}(2010)}]{Mustonen2010-fj}%
  \BibitemOpen
  \bibfield  {author} {\bibinfo {author} {\bibfnamefont {V.}~\bibnamefont
  {Mustonen}}\ and\ \bibinfo {author} {\bibfnamefont {M.}~\bibnamefont
  {Lässig}},\ }\href {\doibase 10.1073/pnas.0907953107} {\bibfield  {journal}
  {\bibinfo  {journal} {Proc. Natl. Acad. Sci. U. S. A.}\ }\textbf {\bibinfo
  {volume} {107}},\ \bibinfo {pages} {4248} (\bibinfo {year}
  {2010})}\BibitemShut {NoStop}%
\bibitem [{\citenamefont {Held}\ \emph {et~al.}(2014)\citenamefont {Held},
  \citenamefont {Nourmohammad},\ and\ \citenamefont {Lässig}}]{Held2014-tn}%
  \BibitemOpen
  \bibfield  {author} {\bibinfo {author} {\bibfnamefont {T.}~\bibnamefont
  {Held}}, \bibinfo {author} {\bibfnamefont {A.}~\bibnamefont {Nourmohammad}},
  \ and\ \bibinfo {author} {\bibfnamefont {M.}~\bibnamefont {Lässig}},\ }\href
  {\doibase 10.1088/1742-5468/2014/09/P09029} {\bibfield  {journal} {\bibinfo
  {journal} {J. Stat. Mech.}\ }\textbf {\bibinfo {volume} {2014}},\ \bibinfo
  {pages} {P09029} (\bibinfo {year} {2014})}\BibitemShut {NoStop}%
\bibitem [{\citenamefont {Kobayashi}\ and\ \citenamefont
  {Sughiyama}(2015)}]{PhysRevLett.115.238102}%
  \BibitemOpen
  \bibfield  {author} {\bibinfo {author} {\bibfnamefont {T.~J.}\ \bibnamefont
  {Kobayashi}}\ and\ \bibinfo {author} {\bibfnamefont {Y.}~\bibnamefont
  {Sughiyama}},\ }\href {\doibase 10.1103/PhysRevLett.115.238102} {\bibfield
  {journal} {\bibinfo  {journal} {Phys. Rev. Lett.}\ }\textbf {\bibinfo
  {volume} {115}},\ \bibinfo {pages} {238102} (\bibinfo {year}
  {2015})}\BibitemShut {NoStop}%
\bibitem [{\citenamefont {Sughiyama}\ and\ \citenamefont
  {Kobayashi}(2017)}]{PhysRevE.95.012131}%
  \BibitemOpen
  \bibfield  {author} {\bibinfo {author} {\bibfnamefont {Y.}~\bibnamefont
  {Sughiyama}}\ and\ \bibinfo {author} {\bibfnamefont {T.~J.}\ \bibnamefont
  {Kobayashi}},\ }\href {\doibase 10.1103/PhysRevE.95.012131} {\bibfield
  {journal} {\bibinfo  {journal} {Phys. Rev. E}\ }\textbf {\bibinfo {volume}
  {95}},\ \bibinfo {pages} {012131} (\bibinfo {year} {2017})}\BibitemShut
  {NoStop}%
\bibitem [{\citenamefont {Nourmohammad}\ \emph {et~al.}(2017)\citenamefont
  {Nourmohammad}, \citenamefont {Rambeau}, \citenamefont {Held}, \citenamefont
  {Kovacova}, \citenamefont {Berg},\ and\ \citenamefont
  {Lässig}}]{Nourmohammad2017-vr}%
  \BibitemOpen
  \bibfield  {author} {\bibinfo {author} {\bibfnamefont {A.}~\bibnamefont
  {Nourmohammad}}, \bibinfo {author} {\bibfnamefont {J.}~\bibnamefont
  {Rambeau}}, \bibinfo {author} {\bibfnamefont {T.}~\bibnamefont {Held}},
  \bibinfo {author} {\bibfnamefont {V.}~\bibnamefont {Kovacova}}, \bibinfo
  {author} {\bibfnamefont {J.}~\bibnamefont {Berg}}, \ and\ \bibinfo {author}
  {\bibfnamefont {M.}~\bibnamefont {Lässig}},\ }\href {\doibase
  10.1016/j.celrep.2017.07.033} {\bibfield  {journal} {\bibinfo  {journal}
  {Cell Rep.}\ }\textbf {\bibinfo {volume} {20}},\ \bibinfo {pages} {1385}
  (\bibinfo {year} {2017})}\BibitemShut {NoStop}%
\bibitem [{\citenamefont {Jarzynski}(1997)}]{Jarzynski1997-ia}%
  \BibitemOpen
  \bibfield  {author} {\bibinfo {author} {\bibfnamefont {C.}~\bibnamefont
  {Jarzynski}},\ }\href {\doibase 10.1103/PhysRevLett.78.2690} {\bibfield
  {journal} {\bibinfo  {journal} {Phys. Rev. Lett.}\ }\textbf {\bibinfo
  {volume} {78}},\ \bibinfo {pages} {2690} (\bibinfo {year}
  {1997})}\BibitemShut {NoStop}%
\bibitem [{\citenamefont {Crooks}(1999)}]{Crooks1999-he}%
  \BibitemOpen
  \bibfield  {author} {\bibinfo {author} {\bibfnamefont {G.~E.}\ \bibnamefont
  {Crooks}},\ }\href {\doibase 10.1103/physreve.60.2721} {\bibfield  {journal}
  {\bibinfo  {journal} {Phys. Rev. E Stat. Phys. Plasmas Fluids Relat.
  Interdiscip. Topics}\ }\textbf {\bibinfo {volume} {60}},\ \bibinfo {pages}
  {2721} (\bibinfo {year} {1999})}\BibitemShut {NoStop}%
\bibitem [{\citenamefont {Seifert}(2012)}]{Seifert2012-dg}%
  \BibitemOpen
  \bibfield  {author} {\bibinfo {author} {\bibfnamefont {U.}~\bibnamefont
  {Seifert}},\ }\href {\doibase 10.1088/0034-4885/75/12/126001} {\bibfield
  {journal} {\bibinfo  {journal} {Rep. Prog. Phys.}\ }\textbf {\bibinfo
  {volume} {75}},\ \bibinfo {pages} {126001} (\bibinfo {year}
  {2012})}\BibitemShut {NoStop}%
\bibitem [{\citenamefont {Sagawa}\ and\ \citenamefont
  {Ueda}(2012)}]{Sagawa2012-ti}%
  \BibitemOpen
  \bibfield  {author} {\bibinfo {author} {\bibfnamefont {T.}~\bibnamefont
  {Sagawa}}\ and\ \bibinfo {author} {\bibfnamefont {M.}~\bibnamefont {Ueda}},\
  }\href {\doibase 10.1103/PhysRevLett.109.180602} {\bibfield  {journal}
  {\bibinfo  {journal} {Phys. Rev. Lett.}\ }\textbf {\bibinfo {volume} {109}},\
  \bibinfo {pages} {180602} (\bibinfo {year} {2012})}\BibitemShut {NoStop}%
\bibitem [{\citenamefont {Sagawa}\ and\ \citenamefont
  {Ueda}(2013)}]{Sagawa2013-in}%
  \BibitemOpen
  \bibfield  {author} {\bibinfo {author} {\bibfnamefont {T.}~\bibnamefont
  {Sagawa}}\ and\ \bibinfo {author} {\bibfnamefont {M.}~\bibnamefont {Ueda}},\
  }\href {\doibase 10.1088/1367-2630/15/12/125012} {\bibfield  {journal}
  {\bibinfo  {journal} {New J. Phys.}\ }\textbf {\bibinfo {volume} {15}},\
  \bibinfo {pages} {125012} (\bibinfo {year} {2013})}\BibitemShut {NoStop}%
\bibitem [{\citenamefont {Barato}\ and\ \citenamefont
  {Seifert}(2014)}]{Barato2014-al}%
  \BibitemOpen
  \bibfield  {author} {\bibinfo {author} {\bibfnamefont {A.~C.}\ \bibnamefont
  {Barato}}\ and\ \bibinfo {author} {\bibfnamefont {U.}~\bibnamefont
  {Seifert}},\ }\href {\doibase 10.1103/PhysRevE.90.042150} {\bibfield
  {journal} {\bibinfo  {journal} {Phys. Rev. E Stat. Nonlin. Soft Matter
  Phys.}\ }\textbf {\bibinfo {volume} {90}},\ \bibinfo {pages} {042150}
  (\bibinfo {year} {2014})}\BibitemShut {NoStop}%
\bibitem [{\citenamefont {Horowitz}\ and\ \citenamefont
  {Esposito}(2014)}]{Horowitz2014-od}%
  \BibitemOpen
  \bibfield  {author} {\bibinfo {author} {\bibfnamefont {J.~M.}\ \bibnamefont
  {Horowitz}}\ and\ \bibinfo {author} {\bibfnamefont {M.}~\bibnamefont
  {Esposito}},\ }\href {\doibase 10.1103/PhysRevX.4.031015} {\bibfield
  {journal} {\bibinfo  {journal} {Phys. Rev. X}\ }\textbf {\bibinfo {volume}
  {4}},\ \bibinfo {pages} {031015} (\bibinfo {year} {2014})}\BibitemShut
  {NoStop}%
\bibitem [{\citenamefont {Kimura}(1968)}]{Kimura1968-nb}%
  \BibitemOpen
  \bibfield  {author} {\bibinfo {author} {\bibfnamefont {M.}~\bibnamefont
  {Kimura}},\ }\href {\doibase 10.1038/217624a0} {\bibfield  {journal}
  {\bibinfo  {journal} {Nature}\ }\textbf {\bibinfo {volume} {217}},\ \bibinfo
  {pages} {624} (\bibinfo {year} {1968})}\BibitemShut {NoStop}%
\bibitem [{\citenamefont {Simmonds}\ \emph {et~al.}(2019)\citenamefont
  {Simmonds}, \citenamefont {Aiewsakun},\ and\ \citenamefont
  {Katzourakis}}]{Simmonds2019-su}%
  \BibitemOpen
  \bibfield  {author} {\bibinfo {author} {\bibfnamefont {P.}~\bibnamefont
  {Simmonds}}, \bibinfo {author} {\bibfnamefont {P.}~\bibnamefont {Aiewsakun}},
  \ and\ \bibinfo {author} {\bibfnamefont {A.}~\bibnamefont {Katzourakis}},\
  }\href {\doibase 10.1038/s41579-018-0120-2} {\bibfield  {journal} {\bibinfo
  {journal} {Nat. Rev. Microbiol.}\ }\textbf {\bibinfo {volume} {17}},\
  \bibinfo {pages} {321} (\bibinfo {year} {2019})}\BibitemShut {NoStop}%
\bibitem [{\citenamefont {Landauer}(1961)}]{Landauer1961-qn}%
  \BibitemOpen
  \bibfield  {author} {\bibinfo {author} {\bibfnamefont {R.}~\bibnamefont
  {Landauer}},\ }\href {\doibase 10.1147/rd.53.0183} {\bibfield  {journal}
  {\bibinfo  {journal} {IBM J. Res. Dev.}\ }\textbf {\bibinfo {volume} {5}},\
  \bibinfo {pages} {183} (\bibinfo {year} {1961})}\BibitemShut {NoStop}%
\bibitem [{\citenamefont {Bull}\ \emph {et~al.}(2007)\citenamefont {Bull},
  \citenamefont {Sanjuán},\ and\ \citenamefont {Wilke}}]{Bull2007-nj}%
  \BibitemOpen
  \bibfield  {author} {\bibinfo {author} {\bibfnamefont {J.~J.}\ \bibnamefont
  {Bull}}, \bibinfo {author} {\bibfnamefont {R.}~\bibnamefont {Sanjuán}}, \
  and\ \bibinfo {author} {\bibfnamefont {C.~O.}\ \bibnamefont {Wilke}},\ }\href
  {\doibase 10.1128/JVI.01624-06} {\bibfield  {journal} {\bibinfo  {journal}
  {J. Virol.}\ }\textbf {\bibinfo {volume} {81}},\ \bibinfo {pages} {2930}
  (\bibinfo {year} {2007})}\BibitemShut {NoStop}%
\bibitem [{\citenamefont {Koelle}\ and\ \citenamefont
  {Rasmussen}(2015)}]{Koelle2015-kg}%
  \BibitemOpen
  \bibfield  {author} {\bibinfo {author} {\bibfnamefont {K.}~\bibnamefont
  {Koelle}}\ and\ \bibinfo {author} {\bibfnamefont {D.~A.}\ \bibnamefont
  {Rasmussen}},\ }\href {\doibase 10.7554/eLife.07361} {\bibfield  {journal}
  {\bibinfo  {journal} {Elife}\ }\textbf {\bibinfo {volume} {4}},\ \bibinfo
  {pages} {e07361} (\bibinfo {year} {2015})}\BibitemShut {NoStop}%
\bibitem [{\citenamefont {Swanstrom}\ and\ \citenamefont
  {Schinazi}(2022)}]{Swanstrom2022-gk}%
  \BibitemOpen
  \bibfield  {author} {\bibinfo {author} {\bibfnamefont {R.}~\bibnamefont
  {Swanstrom}}\ and\ \bibinfo {author} {\bibfnamefont {R.~F.}\ \bibnamefont
  {Schinazi}},\ }\href {\doibase 10.1126/science.abn0048} {\bibfield  {journal}
  {\bibinfo  {journal} {Science}\ }\textbf {\bibinfo {volume} {375}},\ \bibinfo
  {pages} {497} (\bibinfo {year} {2022})}\BibitemShut {NoStop}%
\bibitem [{\citenamefont {Leighton}\ and\ \citenamefont
  {Sivak}(2023)}]{Leighton2023-zi}%
  \BibitemOpen
  \bibfield  {author} {\bibinfo {author} {\bibfnamefont {M.~P.}\ \bibnamefont
  {Leighton}}\ and\ \bibinfo {author} {\bibfnamefont {D.~A.}\ \bibnamefont
  {Sivak}},\ }\href {\doibase 10.1103/PhysRevLett.130.178401} {\bibfield
  {journal} {\bibinfo  {journal} {Phys. Rev. Lett.}\ }\textbf {\bibinfo
  {volume} {130}},\ \bibinfo {pages} {178401} (\bibinfo {year}
  {2023})}\BibitemShut {NoStop}%
\end{thebibliography}
%

\clearpage{}
\onecolumngrid

\appendix
\section{Transition matrix for imbalanced number states for the antibodies and viruses}
\label{Appendix-imbalance}
Because the viral and antibody states can be imbalanced, i.e.,  $n_V \neq n_A$ in general, we distinguish between two types of unbound states. In the case that $n_V > n_A$, there exists states that are inaccessible to the antibodies (i.e., $v=n_A+1,\dots,n_V$), and if the virus transitions to any of them, no change in the antibody state can result in binding, i.e., the virus fully hides from antibodies.  On the other hand, if  $n_A>n_V$, for each viral state there exists more antibody states that do not bind to the virus compared to the balanced case of $n_A=n_V$, resulting in more   evolutionary distraction for antibodies. To reflect these differences, it is useful to introduce three phenotypic states: bound ${\bf B}$, unbound and accessible $\bf U$, and unbound and inaccessible ${\bf \UI}$ (only relevant for the imbalanced case). Defining, $n_- = \min(n_A,n_V)$, and $n_+ = \max(n_A,n_V)$, a system has $n_-$ bound states, $n_-(n_- -1)$ accessible unbound states, and $n_+n_- -n_-^2$ inaccessible unbound states.

Similar to the balanced case, we can invoke symmetry to argue that the diagonal terms of the matrix $W_{a\to a}^{v\to v}$ can be put into three classes, the ones associated with a  bound state  $a=v$, which we denote by $W_{\bf B}$, the ones associated with  accessible unbound states ($a\neq v \, \wedge a,v\leq n_-$), which we denote by $W_{\bf U}$, and those associated with  inaccessible unbound states ($a\neq v \, \wedge\, a | v > n_-$), which we denote by $W_{\UI}$. These three classes of diagonal elements are given by
\EQA
\nonumber W_{\bf B} &=& 1- \left[ (n_V-1)V_+ + (n_A-1)A_-\right],\\
\nonumber W_{\bf U} &=&1- \left[ V_-+A_+ + (n_A-2)A_0+ (n_V-2)V_0\right],\\
 W_{\UI} &=&1- \begin{cases}
V_- + (n_A-1) A_0 + (n_V-2)V_0, &\text{ if } n_v>n_A\\
A_+ + (n_A-2) A_0 + (n_V-1)V_0, &\text{ if } n_A>n_V.
\end{cases}
\EEA
In evaluating these diagonal elements, one needs to account for all the transitions out of the specified state. For example, for $W_\BB$ one should  account for all the transitions out of a bound state.  Since our model allows for only one bound state for a given antibody and virus, i.e., the state $a=v$, all  transitions from a bound state would end in an unbound configuration. There are $(n_V-1)$ such changes through beneficial changes in viruses, each with rate $V_+$, and $(n_A-1)$ such transitions through deleterious changes in antibodies with a rate $A_-$, as reflected in the expression for $W_\BB$. Similarly,   $W_\U$ and $W_{\UI}$ can be calculated by accounting for all the possible transitions, including the neutral transitions to another unbound state (with antibody and viral rates $A_0$ and $V_0$), as shown in the expression for these diagonal terms.

In this general case with an imbalanced number of states ($n_A\neq n_V$), the boundedness of the transition probabilities implies an upper bound for the beneficial substitution probability of viruses, $V_+ \leq \frac{1- (n_A-1)A_-}{n_V-1}$.   Similarly, the upper bound for the   beneficial substitution probability of  antibodies  is $A_+ \leq 1- (V_- + (n_A-2) A_0+ (n_V-2) V_0)$. The neutral substitution probabilities, $A_0$ and $V_0$, scale as $1/N_\text{eff}$  with the effective population size of each system $N_\text{eff}$~\cite{Kimura1968-nb}, which is likely to be much larger than the dimensionality of the subsystems, $n_A$ and $n_V$, yielding $ (n_A-2) A_0+ (n_V-2) V_0 \ll 1$. Given that the deleterious substitution probabilities are exponentially small $V_-\ll 1$~\cite{Kimura1968-nb}, the upper bound on the beneficial substitution probability for antibodies can be approximated as $A_+\lesssim 1$, with no strong dependence on the dimensionality of the system.

\newpage{}
\section{Entropy production and information exchange for a bipartite co-evolutionary system}
\label{appendix-ref}

\noindent {\bf Probability flux.}  We express the probability flux to the state $(a,v)$ by
\EQA
\frac{\d}{\d t} P(a,v) &=&\sum_{a',v'} J_{a'\to a}^{v'\to v}\\
 &=&\sum_{a'\neq a; v'\neq v } { P({a',v'}) W_{a'\to a}^{v'\to v } - P({a,v}) W_{a\to a'}^{v\to v' }}\\
\nonumber &=& \sum_{a'\neq a} \underbrace{ P({a',v}) W_{a'\to a}^{v } -P({a,v}) W_{a\to a'}^{v }}_{J_{a'\to a}^{v}}\\
 &&\qquad {}+ \sum_{v'\neq v}\underbrace{P({a,v'}) W_{a}^{v'\to v} - P({a,v}) W_{a}^{v\to v'}}_{J_{a}^{v'\to v}},
\EEA
where $J_{a'\to a}^{v'\to v}$ is the flux from the state $(a',v')$ to the state $(a,v)$, and $J_{a'\to a}^{v}$ and $J_{a}^{v'\to v}$ are the conditional probability fluxes due to the changes in the antibodies and viruses, respectively. \\
 
\noindent {\bf Change in the  Shannon entropy.}  The Shannon  entropy of the joint system is $S = -\sum_{a,v} P(a,v) \ln P(a,v)$. The change in the Shannon entropy is given by
\EQA
\nonumber\d_t S &=& \d_t\left [- \sum_{a,v} P(a,v) \ln P(a,v)\right]\\
\nonumber &=& - \sum_{a,v} \d_t P(a,v) \ln P(a,v) +\cancelto{0}{ \sum_{a,v} \d t P(a,v) }\\
\nonumber &=& - \sum_{a,v,a',v'} J_{a'\to a}^{v'\to v} \ln P(a,v)  \\
\nonumber &=&- \frac{1}{2}  \sum_{a,v,a',v'} \left (J_{a'\to a}^{v'\to v} - J_{a\to a'}^{v\to v'} \right) \ln P(a,v) \\
\nonumber &=&- \frac{1}{2}  \sum_{a,v,a',v'} J_{a'\to a}^{v'\to v} \left[\ln P(a,v) - \ln P(a',v')\right]\\
 & =& \sum_{a<a', v<v'} J_{a'\to a}^{v'\to v}  \ln \frac{P(a',v')}{P(a,v)},
\label{eq.dTShannon}
\EEA
where we used the fact that probability flux is asymmetric $J_{a'\to a}^{v'\to v} =- J_{a\to a'}^{v\to v'}$.\\

\noindent {\bf Entropy production rate.}  The total entropy production rate $\Phi$ is equal to the sum of the change in the Shannon entropy of the system  $\d_t S$ (eq.~\eqref{eq.dTShannon}) and the total rate of the change in the environment's entropy, which is equal to $\dot S_\text{env.} = \sum_{a<a', v<v'}J_{a'\to a}^{v'\to v} \,\ln\left(W_{a'\to a}^{v'\to v} /W_{a\to a'}^{v\to v'}\right) $. Therefore, the total entropy production rate follows
\EQA
\nonumber\Phi &=&\d_t S +  \dot S_\text{env.} \\
\nonumber&=& \sum_{a<a', v<v'} J_{a'\to a}^{v'\to v}  \ln \frac{P(a',v')}{P(a,v)}+\sum_{a<a', v<v'}J_{a'\to a}^{v'\to v} \ln\frac{W_{a'\to a}^{v'\to v} }{W_{a\to a'}^{v\to v'} }\\
&=&  \sum_{a<a', v<v'} J_{a'\to a}^{v'\to v}  \, \ln \frac{P(a',v') \,W_{a'\to a}^{v'\to v} }{P(a,v) \,W_{a\to a'}^{v\to v'}  }\\
&=&  \underbrace{ \sum_{a<a', v} J_{a'\to a}^{v}  \, \ln \frac{P(a',v) \,W_{a'\to a}^{v} }{P(a,v) \,W_{a\to a'}^{v}  }}_{\Phi^A} + \underbrace{ \sum_{a, v<v'} J_{a}^{v'\to v}  \, \ln \frac{P(a,v') \,W_{a}^{v'\to v} }{P(a,v) \,W_{a}^{v\to v'}  }}_{\Phi^V},
\label{eq.PhiPartial}
\EEA
where we used the bipartite property of the system to define the conditional entropy production rates $\Phi^A$ and $\Phi^V$ due to changes in the antibody and viral ensembles, respectively.\\

We can evaluate the steady-state conditional entropy production rates in our antibody-viral system. Let us start with $\Phi^A$. By inspecting eq.~\eqref{eq.PhiPartial}, we can see that the argument in the sum is symmetric under the state swap $a \leftrightarrow a'$. Given that the sum is avoiding the repetition in the indices and the fact that the probability current  $J$ is only non-zero  between bound and unbound states, we can rewrite the sum in the phenotypic space by conditioning one index to be associated to the bound state and the other index to all the unbound states. Thus, the conditional entropy production $\Phi^A$ associated with antibody changes  is given by
\EQA
\nonumber\Phi^A&=& \sum_{a'\ge a, v }   J_{a' \to a}^{v} \ln{  \frac{W_{a'\to a}^{v }\,\, P(a',v )}{W_{a\to a'}^{v} \,\,P(a,v)}}\\
\nonumber&=&  \sum_{v,a:\BB, a':\U} J_{a':\U \to a:\BB}^{v} \ln{  \frac{W_{a':\U\to a:\BB}^{v }\,\,P(a',v;\U)}{W_{a:\BB\to a':\U}^{v} \,\, P(a,v;\BB)}}\\
&=& n (n-1)\frac{ \sigma}{Z}\,  \ln \frac{A_+ \,\, \pi_\U}{A_- \,\,\pi_\BB},
\EEA
where we have used the stationary state probabilities associated with bound and unbound states $\pi_\BB$, and $\pi_\U$ (eq.~\eqref{eq.Pstat1}).  The prefactor $n(n-1)$ counts for the summation over the $n$ different viral states and the $(n-1)$ possible antibody transitions from bound to unbound states for each of the viral states. 

 With a similar approach, we can evaluate the conditional entropy production rate associated with changes in the viral ensemble, obtaining
\EQA
\nonumber \Phi^V &=&\sum_{a, v' \ge v}   J_{a}^{v'\to v }  \ln{  \frac{W_{ a}^{v' \to v}\,\, P(a,v' )}{W_{a}^{v \to v'} \,\,P(a,v)}} \\
\nonumber&=& \sum_{a,v':\BB, v:\U}  J_{a}^{v':\BB\to v:\U }  \ln{  \frac{W_{ a}^{v':\BB \to v:\U}\,\, \pi_\BB}{W_{a}^{v:\U \to v':\BB} \,\,\pi_\U}} \\
&=&  n (n-1) \frac{\sigma}{Z}\,  \ln \frac{V_+ \,\, \pi_\BB}{V_- \,\,\pi_\U}
\EEA

\noindent {\bf Change in mutual information.} The mutual information flux  for a bipartite system  can  be calculated in the following way:
\EQA
\nonumber \d_t I&=&\d_t  \left[ \sum_{a,v} P(a,v) \ln \frac{P(a,v)}{P(a) P(v)}\right] \\
\nonumber&=& \sum_{a,v} \d_t P(a,v) \ln \frac{P(a,v)}{P(a) P(v)} +\cancelto{0}{ \sum_{a,v} \d_t P(a,v) - \d_t P(a)-\d_t P(v)} \\
\nonumber&=&\sum_{a,v ,a',v'} J_{a'\to a}^{v'\to v} \ln \frac{P(a,v)}{P(a) P(v)}\\
\nonumber &=& \frac{1}{2} \sum_{a,v ,a',v'}  \left( J_{a'\to a}^{v'\to v}- J_{a\to a'}^{v\to v'}\right)\ln \frac{P(a,v)}{P(a) P(v)}\\
\nonumber &=& \frac{1}{2} \sum_{a,v ,a',v'} J_{a'\to a}^{v'\to v} \left( \ln \frac{P(a,v)}{P(a) P(v)}- \ln \frac{P(a',v')}{P(a') P(v')}\right)\\
\nonumber &=& \frac{1}{2} \sum_{a,v ,a'} J_{a'\to a}^{v} \left( \ln \frac{P(a,v)}{P(a) P(v)}- \ln \frac{P(a',v)}{P(a') P(v)}\right) +  \frac{1}{2} \sum_{a,v,v'} J_{a}^{v'\to v} \left( \ln \frac{P(a,v)}{P(a) P(v)}- \ln \frac{P(a,v')}{P(a) P(v')}\right)\\
\nonumber&=&\underbrace{\sum_{a<a', v} J_{a'\to a}^{ v}  \ln \frac{P(v|a)}{P(v|a')}}_{\delta I^A}+  \underbrace{\sum_{v <v',a} J_{ a}^{v'\to v} \ln \frac{P(a|v)}{P(a|v') }}_{\delta I^V}\\
\EEA

\end{document}